\documentclass[aps,graphicx,9pt,onecolumn,superscriptaddress]{revtex4-2}
\usepackage{amsmath,graphicx,cases,amscd}
\usepackage{float,subfigure,multirow}
\usepackage{makecell,blindtext,diagbox,appendix}
\usepackage{color}
\usepackage[colorlinks,linkcolor=blue,urlcolor=blue,anchorcolor=blue,citecolor=blue]{hyperref}
\usepackage{lineno}
\usepackage{easyReview}
\begin{document}
\draft
\title{Revealing inherent quantum interference and entanglement of a Dirac particle}

\author{Wen Ning}
\thanks{These authors contribute equally to this work.}
\affiliation{Fujian Key Laboratory of Quantum Information and Quantum Optics, College of Physics and Information Engineering, Fuzhou University, Fuzhou, Fujian 350108, China}
\author{Ri-Hua Zheng}
\thanks{These authors contribute equally to this work.}
\affiliation{Fujian Key Laboratory of Quantum Information and Quantum Optics, College of Physics and Information Engineering, Fuzhou University, Fuzhou, Fujian 350108, China}
\author{Yan Xia}
\affiliation{Fujian Key Laboratory of Quantum Information and Quantum Optics, College of Physics and Information Engineering, Fuzhou University, Fuzhou, Fujian 350108, China}
\author{Kai Xu}
\affiliation{Institute of Physics, Chinese Academy of Sciences, Beijing 100190, China}
\affiliation{CAS Center for Excellence in Topological Quantum Computation, University of Chinese Academy of Sciences, Beijing 100190, China}
\author{Hekang Li}
\affiliation{Institute of Physics, Chinese Academy of Sciences, Beijing 100190, China}
\author{Dongning Zheng}
\affiliation{Institute of Physics, Chinese Academy of Sciences, Beijing 100190, China}
\affiliation{CAS Center for Excellence in Topological Quantum Computation, University of Chinese Academy of Sciences, Beijing 100190, China}
\author{Heng Fan}
\affiliation{Institute of Physics, Chinese Academy of Sciences, Beijing 100190, China}
\affiliation{CAS Center for Excellence in Topological Quantum Computation, University of Chinese Academy of Sciences, Beijing 100190, China}
\author{Fan Wu}
\email{t21060@fzu.edu.cn}
\affiliation{Fujian Key Laboratory of Quantum Information and Quantum Optics, College of Physics and Information Engineering, Fuzhou University, Fuzhou, Fujian 350108, China}
\author{Zhen-Biao Yang}
\email{zbyang@fzu.edu.cn}
\affiliation{Fujian Key Laboratory of Quantum Information and Quantum Optics, College of Physics and Information Engineering, Fuzhou University, Fuzhou, Fujian 350108, China}
\author{Shi-Biao Zheng}
\email{t96034@fzu.edu.cn}
\affiliation{Fujian Key Laboratory of Quantum Information and Quantum Optics, College of Physics and Information Engineering, Fuzhou University, Fuzhou, Fujian 350108, China}


\begin{abstract}

    Although originally
    predicted in relativistic quantum mechanics, {\sl Zitterbewegung} can also
    appear in some classical systems, which leads to the important question of
    whether {\sl Zitterbewegung} of Dirac particles is underlain by a more
    fundamental and universal interference behavior without classical analogs. We here reveal
    such an interference pattern in phase space, which underlies but goes beyond {\sl %
    Zitterbewegung}, and whose nonclassicality is manifested by the negativity of the phase space quasiprobability distribution, and the associated pseudospin-momentum entanglement. We
    confirm this discovery by numerical simulation and an on-chip experiment,
    where a superconducting qubit and a quantized microwave field respectively
    emulate the internal and external degrees of freedom of a Dirac particle.
    The measured quasiprobability negativities agree well with the numerical
    simulation. Besides being of fundamental importance, the demonstrated nonclassical effects are useful in quantum technology.
\end{abstract}
\vskip0.5cm
\maketitle
\section{Introduction}
The Dirac equation, which describes the wavefunction for a spin-1/2
particle in the framework of relativistic quantum mechanics, represents a
cornerstone of modern physics.
Over the past century, this equation has been producing enduring profound influences
on a wide variety of fields of modern science and technology, ranging from
atomic physics to quantum electrodynamics \cite{Thaller1992}, and from material engineering
\cite{CastroNeto2009,Novoselov2005,Park2008,Tarruell2012} to medical imaging \cite{Beyer1369}.

Despite the fundamental importance of the Dirac equation, the physics
underlying its dynamical solution has not been fully understood owing to the
associated elusive phenomena, exemplified by {\sl Zitterbewegung} ({\sl ZB})
\cite{Sch1930}, the oscillatory motion of a particle, as a result of the
interference between the positive and negative energy components. For a free
electron, the predicted {\sl ZB} has an amplitude on the order of the
Compton wavelength, $\hbar /mc$\ $\sim $\ $10^{-12}$\ m, and thus cannot be
unambiguously observed due to the restriction of the Heisenberg uncertainty
principle. Although whether or not {\sl ZB} really exists in relativistic
quantum mechanics is still an open question \cite{Foldy1950,krekora2004,wang2008,Pedernales2013}, enduring efforts have
been made to its simulations with different quantum systems, including
circuit quantum electrodynamics \cite{Pedernales2013}, ion traps \cite{Lamata2007,Gerritsma2010}, ultracold atoms
\cite{Vaishnav2008,LeBlanc2013,Qu2013,Hasan2022}, semiconductor quantum wells \cite{Schliemann2005,rusin_zitterbewegung_2007,biswas_wave_2014,PhysRevB.72.085217,PhysRevB.73.085323,PhysRevB.77.125303}, graphene \cite{Rusin2009,PhysRevB.74.172305,PhysRevB.76.195439,PhysRevB.75.035305,PhysRevB.78.235321,PhysRevB.78.125419,PhysRevB.91.201402,serna_pseudospin-dependent_2019,lavor_effect_2021,wang_study_2010}, and moir\'{e} excitons \cite{Lavor2021}. These investigations have shed new light on {\sl ZB}, which
itself, however, is not a unique character of Dirac particles as similar
phenomena can also appear in some classical wave systems \cite{Zhang2008,Otterbach2009,Longhi2010,Chen_2019,Dreisow2010,Silva2019}. This leads us to consider whether the {\sl ZB} associated with Dirac particles has a deeper quantum origin that can manifest itself even without {\sl ZB}. Answering this question is critical for understanding the dynamical behaviors of Dirac particles at a more fundamental level, but a deep exploration is still lacking.
\section{Results}
\subsection{Theoretical predictions}
We here present an investigation on this important issue, and unveil a universal quantum interference behavior in the position-momentum space. The nonclassicality of this behavior is manifested by the negativity of the phase space quasiprobability distribution--Wigner function (WF), as well as by the quantum correlation between the spatial and internal degrees of freedom. These quantum signatures distinguish the {\sl ZB} of the Dirac particle, obtained by
integrating the WF over the momentum, from the trembling motion of classical
wavepackets, and more importantly, can express themselves even in the absence
of any negative component. We demonstrate this unique interference pattern
with a circuit, where the spinorial characteristic of a Dirac particle is
encoded in the two lowest energy levels of a superconducting Xmon qubit \cite{barends_superconducting_2014,barends_coherent_2013,song_continuous-variable_2017},
while the position and momentum are mapped to the quadratures of the photonic field. The measured WFs and entanglement entropy agree well with theoretical predictions. Furthermore, we
simulate the Klein tunneling \cite{Klein1929,Dombey1999} in a linear potential field and observe
mesoscopic superpositions of two separated wavepackets in phase space.

We focus on the simplest case that the motion of a Dirac particle is
confined to one dimension (1D), for which the Hamiltonian reduces to%
\begin{equation}\label{eq2}
    H_{\rm D}=c\sigma _{y}\hat{p} + mc^{2}\sigma _{z}.
\end{equation}%
Here $c$ denotes the light speed in the vacuum, $\hat{p}$
represents the momentum operator of the particle with a rest mass of $m$, and $\sigma _{y}$ and $\sigma _{z}$ are the Pauli operators
that endow the Dirac particle a spinor characteristic, manifested by a
two-component wavefunction, where the spatial position and momentum are
correlated with the degree of freedom defined in an ``internal space", which
will be referred to as pseudospin for simplicity.
Unlike the Schr\"{o}dinger equation, the Dirac equation is linear in both the
time- and space-derivatives, satisfying the Lorentz-covariance, and includes
the spin degree of freedom at the {\sl ab initio} level by describing the
wave function in terms of a spinor. These features have led to remarkable accomplishments, including predictions of the spin-1/2 feature of electrons and the existence of anti-particles indicated by the negative-energy component accompanying the positive one, and introduction of the spin-orbit interaction that led to a more refined fine structure description of the spectrum. These predictions are based on stationary solutions of
the Dirac equation and show excellent agreements with experiments.
As the Hamiltonian commutes with the
momentum operator, it is illuminating to uncover the physics in the momentum
representation, where the momentum operator $\hat{p}$ can be
taken as a parameter $p$. For a specific value of $p$, $H_{\rm D}$ has two
eigenvalues $\pm E_{p}=\pm \sqrt{p^{2}c^{2}+m^{2}c^{4}}$, with the
corresponding eigenstates $\left\vert \phi _{+}(p)\right\rangle =\left( \cos
\phi _{p}, {\rm i}\sin \phi _{p}\right) ^{T}$ and $\left\vert \phi
_{-}(p)\right\rangle =\left( {\rm i}\sin \phi _{p},\cos \phi _{p}\right) ^{T}$,
where $\tan (2\phi _{p})=\frac{p}{mc}$. Suppose that the system is initially
in the product state
\begin{equation}
    \left\vert \psi (0)\right\rangle =\int \mathrm{d}p\ \xi _{p}\left\vert p\right\rangle
    \left\vert X\right\rangle ,
\end{equation}%
where $\left\vert \pm X\right\rangle =\frac{1}{\sqrt{2}}\left( 1,\pm
1\right) ^{T}$ and $\xi _{p}$ denotes the wave function in the momentum
representation. Under the Dirac Hamiltonian, the system evolves as%
\begin{equation}
    \left\vert \psi (t)\right\rangle =\int \mathrm{d}p\ \left\vert p\right\rangle \xi
    _{p}(\cos \varphi _{t}\left\vert X\right\rangle -{\rm i}{\rm e}^{-2{\rm i}\phi _{p}}\sin
    \varphi _{t}\left\vert -X\right\rangle ),
\end{equation}%
where $\varphi _{t}=E_{p}t/\hbar $. This directly yields the average
position evolution,%
\begin{eqnarray}
    \left\langle x(t)\right\rangle &=&\left\langle x(0)\right\rangle
    +\left\langle v(0)\right\rangle t \\
    &&+\hbar \int \mathrm{d}p\ \left\vert \xi _{p}\right\vert ^{2}x_{p}(1-\cos 2\varphi
    _{t}),  \nonumber
\end{eqnarray}%
where $x_{p}=\mathrm{d}\phi _{p}/\mathrm{d}p$, $\left\langle x(0)\right\rangle $ represents
the average value of the initial position, and $\left\langle
v(0)\right\rangle $ denotes the initial mean velocity. The {\sl ZB}, manifested by
the last term, is observable only in the intermediate regime where $mc$ is
comparable with $p$.

In the non-relativistic regime $\left|p\right|\ll mc$, the {\sl ZB} amplitude $A\simeq \lambda_{c}/2$, where $\lambda _{c}=\hbar /mc$ is the Compton wavelength, which sets the lower bound for the uncertainty of the position, and consequently, the {\sl ZB} cannot be observed. In the far-relativistic regime $\left\vert p\right\vert \gg mc$, $A\ll \hbar /2\left\langle p\right\rangle \ll \delta x$, where $\delta x=\hbar /2\delta p$ is the limitation of precision attainable for any position measurement imposed by the Heisenberg uncertainty relation.

As we have noted, {\sl ZB} itself does not manifest quantum effects, but is closely related to quantum entanglement between the internal and spatial degrees of freedom, produced by their coupling.
Under the time evolution, the populations of two states
$\left\vert \pm X\right\rangle $\ become
increasingly balanced, and the entropy tends to $1$. Due to this
entanglement, the spatial quantum interference appears when the WF is
correlated with the projection of the pseudospin along some basis, e.g., \{$%
\left\vert \pm B\right\rangle $\}. The WFs associated to $\left\vert \pm
B\right\rangle $\ are respectively
\begin{equation}
    {\cal W}_{\pm }(x,p)=\frac{1}{{\rm \pi} \hbar }\int \mathrm{d}v\ \phi _{\pm }^{\ast
    }(p+v)\phi _{\pm }(p-v)e^{-2{\rm i}vx/\hbar },
\end{equation}%
where $\phi _{\pm }(p)=\xi _{p}\left\langle \pm B\right\vert (\cos \varphi
_{t}\left\vert X\right\rangle -{\rm i}{\rm e}^{-2{\rm i}\phi _{p}}\sin \varphi _{t}\left\vert
-X\right\rangle )$. During the evolution, the wavepacket is continually
deformed under the competition between the momentum-dependent and static
energy terms in the Dirac Hamiltonian, which leads to a nonlinear dependence
of the energy on the momentum. This nonlinear process evolves an initial
Gaussian wavepacket to a non-Gaussian one, manifesting pseudospin-dependent
quantum interference signatures. The {\sl ZB} phenomenon appears as the
integral of the weighted mixture of the two WFs over the momentum, which
reflects the classical probability distribution, but does not manifest the
underlying quantum nature. It should be noted that the presence of {\sl ZB} is challenged by the claim that the positive and
negative components could not be assigned to a single particle \cite{Foldy1950}, however, recent experimental
evidence indicates nature does not prohibit the existence of a quantum
superposition of a particle with its antiparticle \cite{Aaij2021}. We further note that
even when the particle remains in the positive branch, there still exists
phase space quantum interference, though {\sl ZB} disappears.
\begin{figure}
    \includegraphics[width=8cm]{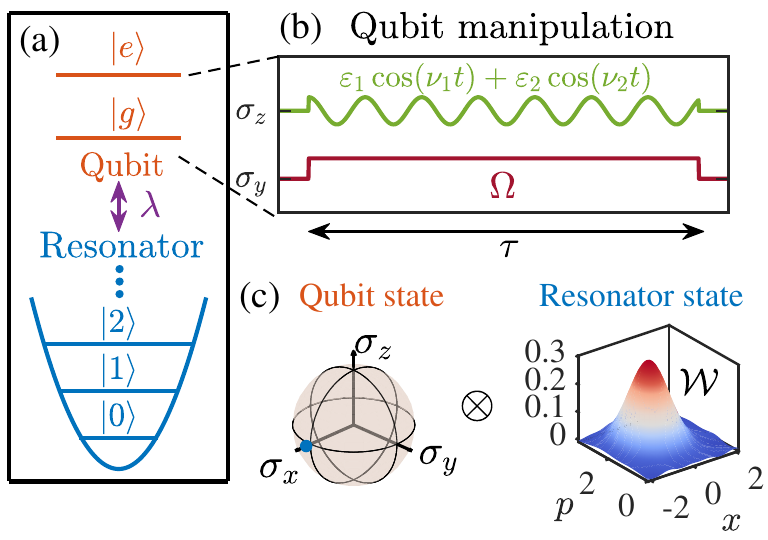}
    \caption{Diagram and pulse for simulating 1D Dirac
        particles. (a) Analog of the Dirac particle in a circuit. The internal and
        external degrees of freedom of the Dirac particle are respectively encoded
        in a superconducting qubit and the field mode in a microwave resonator,
        whose quadratures behave like the position ($x$) and momentum ($p$) of the
        particle. (b) Engineering of the Dirac Hamiltonian. The coupling between the
        internal and external states of the spinor is simulated by transversely
        driving the qubit with continuous microwave, and longitudinally modulating
        it with two alternating current (AC) fluxes. The first parametric modulation controls the
        qubit-resonator interaction at the second upper sideband, which together
        with the transverse driving, effectively realizes the momentum term. The
        second parametric modulation serves for adjusting the effective mass of the
        simulated particle. (c) Pictorial representation of the system's initial
        state. The qubit's state $\left\vert X\right\rangle $ is represented by its
        Bloch vector, while the resonator's coherent state $\left\vert {\rm i}\sqrt{2}%
        \right\rangle $ is characterized by its WF.}
    \label{fig1}
\end{figure}
\subsection{Device and experimental scheme}
The simulation is performed with a superconducting qubit of angular frequency $\omega_0$ that encodes the internal state of the simulated spinor, whose position and momentum are mapped onto the two quadratures of the microwave field stored in a bus resonator, defined as $\hat{x}=\frac{1}{\sqrt{2}}(a^{\dagger }+a)$ and $\hat{p}=\frac{{\rm i}}{\sqrt{2}}(a^{\dagger }-a)$, where $a^{\dagger }$ and $a$ denote the creation and annihilation operators for the photonic field of angular frequency $\omega_{r}$. If we take $\hbar =1$,
$\hat{x}$ and $\hat{p}$ satisfy the same commutation relation as the position and momentum operators. The qubit is subjected to two longitudinal parametric modulations with amplitudes $\varepsilon _{j}$ and angular frequencies $\nu _{j}$ ($j=1,2$), and a transverse continuous microwave driving with an amplitude $\Omega $ (Fig. \ref{fig1}(a)). With the choice $\omega _{r}=\omega_{0}+2\nu _{1}$, the resonator is coupled to the qubit at the second upper sideband of the first modulation with the effective strength $\eta =\lambda J_{2}(\mu )/2$ \cite{Zheng2022}, where $J_{n}(\mu )$ is the $n$th Bessel function of the first kind, with $\mu =\varepsilon
_{1}/\nu _{1}$. When $\Omega \gg 2\eta $, the transverse drive effectively transforms the rotating-wave interaction into an equal combination of rotating- and counter-rotating-wave interactions, simulating the coupling between the internal and external degrees of freedom of the spinor. Then the resulting effective Hamiltonian reduces to the form of Eq. (\ref{eq2}), with the correspondences
$\sigma_{y}={\rm i}\vert g\rangle\langle e \vert - {\rm i}\vert e\rangle \langle g \vert$, $\sigma_{z}=\vert e\rangle\langle e \vert - \vert g\rangle \langle g \vert$, $c^{\ast }=\sqrt{2}%
\eta $ and $m^{\ast }c^{\ast 2}=\varepsilon _{2}/4$, where $c^{\ast }$ and $m^{\ast }$ denote the effective light speed and mass of the Dirac particle in the simulation, respectively (see Supplementary Note 1).

Before the experiment, both the resonator and the spinor qubit are
initialized to their ground states. The experiment starts with the application of
a pulse to the resonator, translating its state along the $p$-axis in phase
space by an amount of $p_{0}=2$, and thus transforming the initial vacuum
state to the coherent state $\left\vert {\rm i}\sqrt{2}\right\rangle $. Then a $%
{\rm \pi} /2$ rotation is performed on the test qubit, transforming it from $%
\left\vert g\right\rangle $ to $\left\vert X\right\rangle $ at the operating
frequency $\omega _{0}/(2{\rm \pi})=5.26$ GHz. The initial qubit-resonator state is
pictorially shown in Fig. \ref{fig1}(b). After this initial state preparation, two
parametric modulations with frequencies $\nu _{1}/(2{\rm \pi})=160$ MHz and $\nu _{2}/(2{\rm \pi})=33.4$ MHz are applied to the qubit. These modulations, together with the
transverse microwave driving at the frequency $\omega _{0}$, couple the
qubit to the bus resonator with a fixed frequency of $\omega _{r}/(2{\rm \pi})=5.584$ GHz,
effectively realizing the Dirac Hamiltonian in the rotating frame. The ratio
between the effective momentum and mass of the Dirac particle is controlled by
adjusting the modulation amplitudes $\varepsilon _{1}$ and/or $\varepsilon
_{2}$. Detailed system parameters are shown in the Supplementary Table 1.
\begin{figure}
    \includegraphics[width=8.5cm]{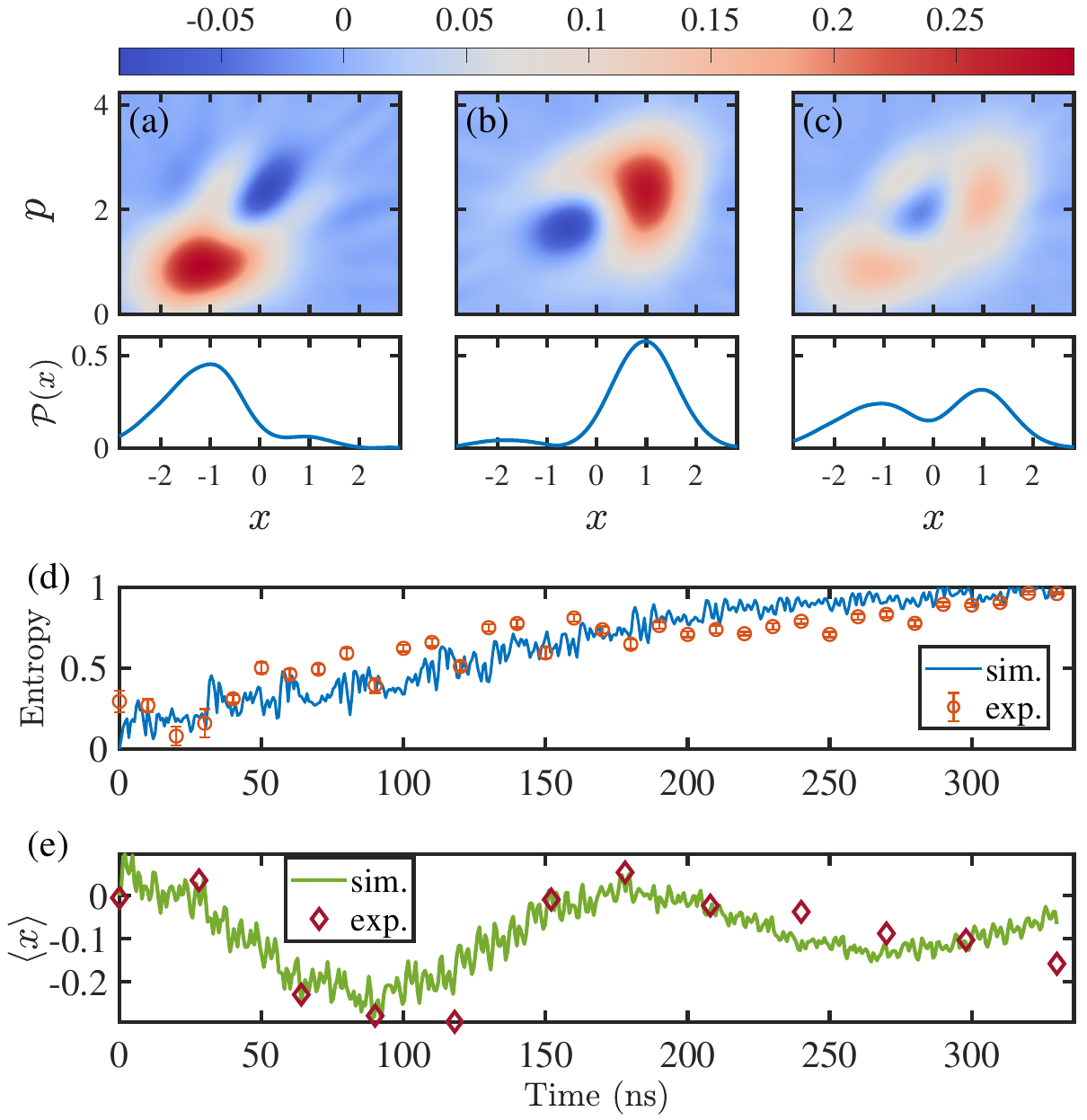}
    \caption{Observation of phase space quantum interference. (a),
        (b) WFs correlated with the basis states $\left\vert g\right\rangle $ (a)
        and $\left\vert e\right\rangle $ (b) of the test qubit; (c) WF irrespective
        of the test qubit's state, all measured after an evolution time of 330 ns.
        Probability distributions ${\cal P}(x)$ with respect to the quadrature $x$,
        shown in the lower panels, are obtained by integrating the WF ${\cal W}(x,p)$
        over $p$. (d) Entropy evolution of the test qubit. This entropy is
        directly obtained from the density matrix of the qubit, measured
        irrespective of the resonator's state. Symbols ``exp." and ``sim." represent the experimental and corresponding simulated data, respectively.
        (e) Evolution of the average value of the resonator's quadrature $\left\langle x\right\rangle $. The value at each point is extracted from the
        measured WF. }
    \label{fig2}
\end{figure}
\subsection{{\sl ZB} interference behaviors}
We investigate the interference behaviors with the choice of effective spinor frequency $\omega = m^{*} c^{*}= \sqrt{2}\eta p_{0}$, with $c^{*} = \sqrt{2}\eta$, which ensures that the Dirac particle described by the Dirac equation $H_{\rm D}$ has an initial momentum of $p_0$.
After a preset evolution time, both the
parametric modulations and microwave driving are switched off. This is
followed by Wigner tomography, realized by performing a phase space
displacement, $D(\gamma )={\rm e}^{\gamma a^{\dagger }-\gamma ^{\ast }a}$, to the
resonator and then tuning an ancilla qubit on resonance with the resonator.
The photon number population of the displaced resonator field, ${\cal P}%
_{n}(\gamma )$, inferred from the measured Rabi oscillation signals,
directly yields the WF, %
${\cal W}(x,p)=\frac{1}{{\rm \pi} }\mathrel{\mathop{\sum }\limits_{n}}(-1)^{n}{\cal P}_{n}(\gamma )$ \cite{Zheng2022,Hofheinz2009},
where $x=\sqrt{2}\mathop{\rm Re}\gamma $ and $p=\sqrt{2}\mathop{\rm Im}\gamma $.
The nonclassical features of the simulated Dirac particle can be
revealed by the WFs of the resonator conditional on the detection of the
qubit state. The WFs correlated with the measurement outcomes $\left\vert
g\right\rangle $ and $\left\vert e\right\rangle $ are presented in the upper
panels of Figs. \ref{fig2}(a) and \ref{fig2}(b), and the result irrespective of the test qubit's
state is displayed in Fig. \ref{fig2}(c), all measured after an evolution time $330$ ns. As expected, during the evolution the initial Gaussian wavepacket
is split into two parts, propagating towards opposite directions. The WF
associated with each qubit state displays a region of negativity that is a
purely quantum-mechanical effect. This result can be interpreted as follows.
Under the Dirac Hamiltonian, each component accumulates a phase that
nonlinearly depends on the ``momentum" and ``mass" as $\sqrt{p^{2}+(m^{\ast
    }c^{\ast })^{2}}c^{\ast }t$. Such a process corresponds to a non-Gaussian
operation, turning a Gaussian state to a non-Gaussian state. The two rods
sprouting from the bulk of the distorted wavepacket interfere with each
other, resulting in a negative quasiprobability distribution in the region
between them. The lower panels show the probability distribution ${\cal P}%
(x) $ with respect to the quadrature $x$, obtained by integrating ${\cal W}%
(x,p)$ over $p$.

Another important feature associated with the simulated particle is the
production of quantum entanglement between its internal and external degrees
of freedom. To quantitatively characterize the behavior, we present the von
Neumann entropy of the test qubit, ${\cal S}=-{\rm tr}(\rho _{q}\log _{2}\rho_{q}) $, measured for different evolution time $t$ in Fig. \ref{fig2}(d), where $\rho
_{q}$ denotes the reduced density operator of this qubit. The measured results (red circles) agree with the numerical simulation (blue curve), where the small fluctuations are due to the fast
Rabi oscillations. {\sl ZB} is manifested in the time-evolving mean position, which is related to ${\cal P}(x)$ by $\langle x \rangle = {\rm tr}(\rho x)$, where $\rho$ is the corresponding density matrix deduced by the measured WF data (see Supplementary Note 5). This evolution, inferred from the measured WF, is
presented in Fig. \ref{fig2}(e), which coincides with the simulation (green curve),
confirming that {\sl ZB} has a deeper root that is of pure quantum
characteristic.
We notice that the numerical simulation curves of Figs. \ref{fig2}(d) and \ref{fig2}(e) both have high frequency vibrations. The high-frequency signals mainly come from the first parametric modulation with frequency $2{\rm \pi}\times160$ MHz, whose Jacobi-Anger expansion will produce two major frequency components of  $2{\rm \pi}\times160$ MHz and $2{\rm \pi}\times320$ MHz around the dynamically resonant frequency (see Supplementary Note 1).

\begin{figure}
\includegraphics[width=8.5cm]{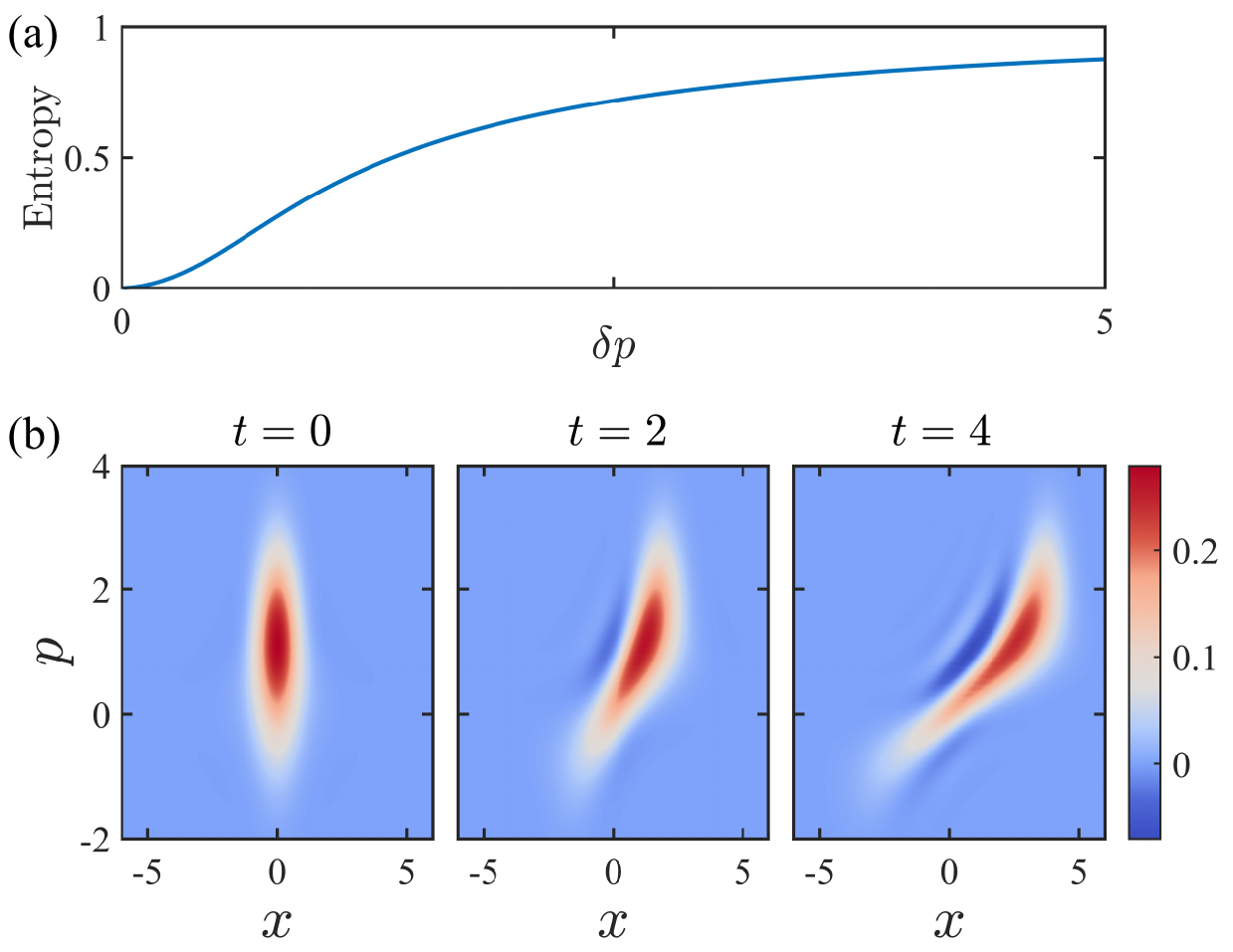}
\caption{Quantum characteristics in the positive branch. (a) Entanglement entropy versus the momentum spread. (b) WF evolution with a momentum spread $\delta p=1$. In the numerical simulations, the Dirac particle is supposed to have a mass of $m=1$ and a Gaussian momentum distribution with a mean value of $p_0=1$. For simplicity, we here take the natural unit with $\hbar=c=1$.}
\label{fig3n}
\end{figure}

Although {\sl ZB} appears only when the system is in a superposition of
positive and negative components, phase space quantum interference is
actually a universal inherent characteristic of Dirac particles, which can
manifest itself even without negative components. This point can be
illustrated with the representative example, where the momentum has a
Gaussian distribution, centered at $p_{0}$ with the spread $\delta p$. When
restricted to the positive branch, the system state can be written as
\begin{equation}
    \left\vert \psi (t)\right\rangle =\int \mathrm{d}p\ {\rm e}^{{\rm i}\theta _{p}(t)}\xi
    _{p}\left\vert p\right\rangle \left\vert \phi _{+}(p)\right\rangle ,
\end{equation}
where $\xi_{p}=(\delta p \sqrt{2{\rm \pi}} )^{-1/2}{\rm e}^{-(p-p_{0})^{2}/(2\delta p)^2} $. This implies that the internal degrees freedom is necessarily
entangled with the momentum except for a plane wave with $\delta
p\rightarrow 0$. As it is experimentally difficult to prepare such an entangled state, we reveal the associated quantum feature by numerical simulation. The entanglement entropy as a function of $\delta_p$ is shown in Fig. \ref{fig3n}(a), which is independent of the evolution time. We here have set $\hbar=c=1$ and $m=p_0=1$. Unlike the entropy, the phase space interference pattern is time-dependent. To clearly reveal such an interference behavior, we assume that $\delta p=1$, and $\theta _{p}(0)=0$. The unconditional WFs, for different evolution times under the ideal Dirac Hamiltonian, are shown in Fig. \ref{fig3n}(b). Unexpectedly, the WF displays a time-evolving quantum interference pattern, even without correlating the result with the internal state. We note that the phase space quantum interference effects were previously predicted for some special states with both positive and negative components \cite{Pedernales2013}, but the presence of such effects without {\sl ZB} has not been revealed. The WFs correlated with $\vert g\rangle$ and $\vert e\rangle$, together with the evolutions of mean position and entanglement entropy, are displayed in the Supplementary Note 3.
\subsection{Klein tunneling}
\begin{figure}
    \includegraphics[width=8.4cm]{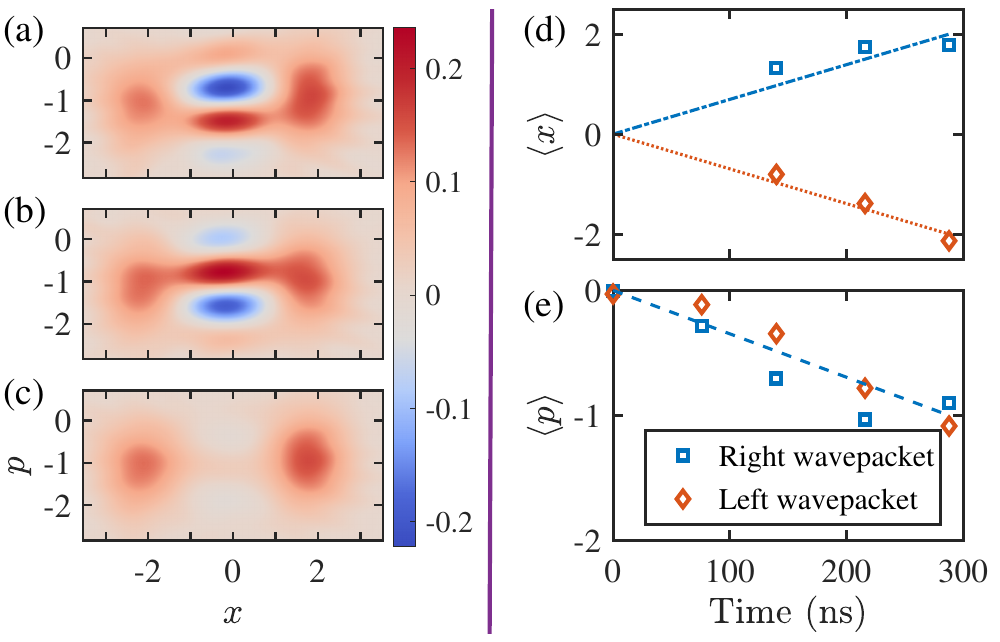}
    \caption{Simulation of Klein's tunneling. The potential is
        simulated by applying a drive to the resonator with the strength $\epsilon
        =2{\rm \pi} \times 0.39$ MHz. The simulation is performed without the second parametric modulation
        so that the system dynamics are analogous to a massless particle penetrating a
        linear potential. (a) Measured WF of the resonator conditional on the
        detection of the spinorial state $\left\vert g\right\rangle $. (b)
        Resonator's WF correlated with the detection of $\left\vert e\right\rangle $.
        (c) Unconditional WF. These WFs are reconstructed for the system evolving
        from the initial ground state $\left\vert g,0\right\rangle $ for a time $288$ ns.
        (d), (e) Evolutions of the measured $\left\langle x\right\rangle $ and $%
        \left\langle p\right\rangle $. The dots and diamonds denote the measured
        values for wavepackets along the positive and negative directions of the $x$-axis,
        and the triangles correspond to the results without discrimination between
        the moving directions.}
    \label{fig3}
\end{figure}

Pushing one step further, we simulate Klein tunneling in a linear
potential field \cite{Silva2019,Klein1929}. It was first noted by Klein that a relativistic
electron may exhibit a counter-intuitive behavior when confronted with a
semi-infinite step potential with $V=0$ and $V_{0}$ for $x<0$ and $x\geq 0$,
respectively. This occurs in the regime $V_{0}>E+mc^{2}$, for which the
electron can propagate through the barrier without damping, where it is
transformed into a positron, where $E$ denotes the initial kinetic energy.
In our setup, it is not easy to engineer the step-shaped potential. However,
a linear potential can be added to the Dirac Hamiltonian {\sl in situ} by
applying a continuous microwave to the resonator, given by $V=\sqrt{2}%
~\epsilon x$, where $\epsilon $ is the amplitude of the drive, and
set to be $2{\rm \pi} \times 0.39$ MHz in our experiment. For simplicity, the simulation is performed
for the choice $\varepsilon_{2}=0$. Figures \ref{fig3}(a) and \ref{fig3}(b) showcase the WFs of the resonator
correlated with the states $\left\vert g\right\rangle $ and $\left\vert
e\right\rangle $ of the spinor qubit, respectively, and Fig. \ref{fig3}(c) presents the
result irrespective of the qubit's state, all measured after an evolving
time $t=288$ ns. As expected, the linear potential drags the phase space
evolution down along the $p$-axis by an amount $\sqrt{2}\epsilon t$, but
does not affect the motion along the $x$-axis. The resulting cat-like state
was previously predicted to exist as a solution of a relativistic spin-1/2
charged particle in an external magnetic field \cite{Bermudez2007}, but has not been
characterized in previous simulations \cite{Gerritsma2011,Salger2011}.
To illustrate this phenomenon more clearly, we display the time evolutions
of the measured $\left\langle x\right\rangle $ and $\left\langle
p\right\rangle $ in Figs. \ref{fig3}(d) and \ref{fig3}(e), respectively, where the dots and
diamonds respectively denote the results for wavepackets along the positive
and negative directions of the $x$-axis. The measured results imply that the two
wavepackets have the same momentum at each moment, but move along the
opposite directions. This can be explained as follows. For $x>0$, the
momentum of the Dirac particle is given by $p^{2}=(E-V)^{2}-m^{2}$ ($c=1$),
with the group velocity $v_{g}=\mathrm{d}E/\mathrm{d}p=p/(E-V)$ \cite{Dombey1999}. When the massless
particle moves from left to right with $E<V$, $p$ is assigned with its
negative solution, so that $v_{g}$ is positive.
\section{Discussion}
We have performed an investigation on the dynamical evolution
of the Dirac particle, showing that the competition between its dynamic and
static energies leads to a time-evolving phase space quasi-probability
distribution, which underlies its spatial motion. The quantum interference
signatures appear in the phase space spanned by position and momentum, but disappear when the momentum is traced out. We
demonstrate this discovery with numerical simulations and with a circuit
experiment, where a superconducting qubit represents the internal degree of
freedom, which is coupled to the microwave field in a resonator that encodes
the spatial degree of freedom. The measured negativities of the WFs and
entanglement entropy distinguish the {\sl ZB} of the Dirac particle from
that exhibited by the classical systems.
In addition to fundamental interest, the
demonstrated nonclassicality can serve as a resource for quantum-enhanced
sensing \cite{Hyukjoon2019}.

\section{DATA AVAILABILITY}
All data needed to evaluate the conclusions in the paper are present in the paper
and the Supplementary Materials. Additional data related to this paper may be
requested from the authors.

\section{Acknowledgments}
This work was supported by the National Natural Science Foundation of China (Grants No. 12274080, No. 11875108, No. 12204105, No. 11934018, No. 92065114, and No. T2121001), Innovation Program for Quantum Science and Technology (Grant No. 2021ZD0300200), the Strategic Priority Research Program of Chinese Academy of Sciences (Grant No. XDB28000000), the Key-Area Research and Development Program of Guangdong Province, China (Gran No. 2020B0303030001), Beijing Natural Science Foundation (Grant No. Z200009), the Natural Science Funds for Distinguished Young Scholar of Fujian Province under Grant 2020J06011, and Project from Fuzhou University under Grants No. JG202001-2 and No. 049050011050.
\section{Author Contributions}
S.-B.Z. theoretically predicted phase space quantum interference of Dirac particles and conceived the experimental simulation.
W.N. and R.-H.Z. performed numerical and experimental simulations, under the supervision of Z.-B.Y. and S.-B.Z., F.W. analyzed the data with support from
R.-H.Z., S.-B.Z., Z.-B.Y., and F.W. co-wrote the paper. All authors contributed to the interpretation of observed phenomena and helped to improve the presentation of the paper.
\section{COMPETING INTERESTS}
The authors declare no competing financial or non-financial interests.
\section{References}
%

\section{Figure Legends}
\subsection*{Figure 1}
Diagram and pulse for simulating 1D Dirac
particles. (a) Analog of the Dirac particle in a circuit. The internal and
external degrees of freedom of the Dirac particle are respectively encoded
in a superconducting qubit and the field mode in a microwave resonator,
whose quadratures behave like the position ($x$) and momentum ($p$) of the
particle. (b) Engineering of the Dirac Hamiltonian. The coupling between the
internal and external states of the spinor is simulated by transversely
driving the qubit with continuous microwave, and longitudinally modulating
it with two alternating current (AC) fluxes. The first parametric modulation controls the
qubit-resonator interaction at the second upper sideband, which together
with the transverse driving, effectively realizes the momentum term. The
second parametric modulation serves for adjusting the effective mass of the
simulated particle. (c) Pictorial representation of the system's initial
state. The qubit's state $\left\vert X\right\rangle $ is represented by its
Bloch vector, while the resonator's coherent state $\left\vert {\rm i}\sqrt{2}%
\right\rangle $ is characterized by its WF.
\subsection*{Figure 2}
Observation of phase space quantum interference. (a),
(b) WFs correlated with the basis states $\left\vert g\right\rangle $ (a)
and $\left\vert e\right\rangle $ (b) of the test qubit; (c) WF irrespective
of the test qubit's state, all measured after an evolution time of 330 ns.
Probability distributions ${\cal P}(x)$ with respect to the quadrature $x$,
shown in the lower panels, are obtained by integrating the WF ${\cal W}(x,p)$
over $p$. (d) Entropy evolution of the test qubit. This entropy is
directly obtained from the density matrix of the qubit, measured
irrespective of the resonator's state. Symbols ``exp." and ``sim." represent the experimental and corresponding simulated data, respectively.
(e) Evolution of the average value of the resonator's quadrature $\left\langle x\right\rangle $. The value at each point is extracted from the
measured WF.
\subsection*{Figure 3}
Quantum characteristics in the positive branch. (a) Entanglement entropy versus the momentum spread. (b) WF evolution with a momentum spread $\delta p=1$. In the numerical simulations, the Dirac particle is supposed to have a mass of $m=1$ and a Gaussian momentum distribution with a mean value of $p_0=1$. For simplicity, we here take the natural unit with $\hbar=c=1$.
\subsection*{Figure 4}
Simulation of Klein's tunneling. The potential is
simulated by applying a drive to the resonator with the strength $\epsilon
=2{\rm \pi} \times 0.39$ MHz. The simulation is performed without the second parametric modulation
so that the system dynamics are analogous to a massless particle penetrating a
linear potential. (a) Measured WF of the resonator conditional on the
detection of the spinorial state $\left\vert g\right\rangle $. (b)
Resonator's WF correlated with the detection of $\left\vert e\right\rangle $.
(c) Unconditional WF. These WFs are reconstructed for the system evolving
from the initial ground state $\left\vert g,0\right\rangle $ for a time $288$ ns.
(d), (e) Evolutions of the measured $\left\langle x\right\rangle $ and $%
\left\langle p\right\rangle $. The dots and diamonds denote the measured
values for wavepackets along the positive and negative directions of the $x$-axis,
and the triangles correspond to the results without discrimination between
the moving directions.
\end{document}


\title{Supplementary Material for\\ ``Revealing inherent quantum interference and entanglement of a Dirac particle''}

\author{Wen Ning}
\thanks{These authors contribute equally to this work.}
\affiliation{Fujian Key Laboratory of Quantum Information and Quantum Optics, College of Physics and Information Engineering, Fuzhou University, Fuzhou, Fujian 350108, China}
\author{Ri-Hua Zheng}
\thanks{These authors contribute equally to this work.}
\affiliation{Fujian Key Laboratory of Quantum Information and Quantum Optics, College of Physics and Information Engineering, Fuzhou University, Fuzhou, Fujian 350108, China}
\author{Yan Xia}
\affiliation{Fujian Key Laboratory of Quantum Information and Quantum Optics, College of Physics and Information Engineering, Fuzhou University, Fuzhou, Fujian 350108, China}
\author{Kai Xu}
\affiliation{Institute of Physics, Chinese Academy of Sciences, Beijing 100190, China}
\affiliation{CAS Center for Excellence in Topological Quantum Computation, University of Chinese Academy of Sciences, Beijing 100190, China}
\author{Hekang Li}
\affiliation{Institute of Physics, Chinese Academy of Sciences, Beijing 100190, China}
\author{Dongning Zheng}
\affiliation{Institute of Physics, Chinese Academy of Sciences, Beijing 100190, China}
\affiliation{CAS Center for Excellence in Topological Quantum Computation, University of Chinese Academy of Sciences, Beijing 100190, China}
\author{Heng Fan}
\affiliation{Institute of Physics, Chinese Academy of Sciences, Beijing 100190, China}
\affiliation{CAS Center for Excellence in Topological Quantum Computation, University of Chinese Academy of Sciences, Beijing 100190, China}
\author{Fan Wu}
\email{t21060@fzu.edu.cn}
\affiliation{Fujian Key Laboratory of Quantum Information and Quantum Optics, College of Physics and Information Engineering, Fuzhou University, Fuzhou, Fujian 350108, China}
\author{Zhen-Biao Yang}
\email{zbyang@fzu.edu.cn}
\affiliation{Fujian Key Laboratory of Quantum Information and Quantum Optics, College of Physics and Information Engineering, Fuzhou University, Fuzhou, Fujian 350108, China}
\author{Shi-Biao Zheng}
\email{t96034@fzu.edu.cn}
\affiliation{Fujian Key Laboratory of Quantum Information and Quantum Optics, College of Physics and Information Engineering, Fuzhou University, Fuzhou, Fujian 350108, China}

\maketitle

\section*{Supplementary Note 1: Synthesis of the Dirac Hamiltonian}


To realize the spinor-momentum coupling, the excitation energy of the test qubit is periodically modulated as ($\hbar =1$ hereafter)
\begin{eqnarray}
\omega _{\rm q}(t)=\omega _{0}+\varepsilon _{1}\cos (\nu _{1}t)+\varepsilon
_{2}\cos (\nu _{2}t),
\end{eqnarray}%
where $\omega _{0}$ is the mean excitation energy, and $\varepsilon _{j}$ and $\nu _{j}$ ($j=1,2$) are the corresponding modulation amplitude and angular frequency, respectively. In addition to interaction with the resonator, the qubit is driven by a continuous microwave with frequency $\omega _{0}$. The system Hamiltonian is given by
\begin{eqnarray}\label{full}
H=H_{0}+H_{\rm I},
\end{eqnarray}
where%
\begin{eqnarray}
\begin{split}
        H_{0} &=\omega _{\rm r}a^{\dagger }a+[\omega _{0}+\varepsilon _{1}\cos (\nu
    _{1}t)]\left\vert e\right\rangle \left\langle e\right\vert , \\
    H_{\rm I} &=\varepsilon _{2}\cos (\nu _{2}t)\left\vert e\right\rangle
    \left\langle e\right\vert +(-\lambda a^{\dagger }\left\vert g\right\rangle
    \left\langle e\right\vert +\Omega {\rm e}^{{\rm i}\theta }{\rm e}^{{\rm i}\omega _{0}t}\left\vert
    g\right\rangle \left\langle e\right\vert +\rm{H.c.}),
\end{split}
\end{eqnarray}%
$\lambda $ is the qubit-resonator coupling strength, and $\Omega $ ($\theta $) denotes the amplitude (phase) of the classical driving field. We here assume $\omega _{\rm r}=\omega _{0}+2\nu _{1}$. With this setting, the resonator interacts with the qubit at the second sideband associated with the first modulation, while the microwave drive works at the carrier. Performing the transformation
\begin{eqnarray}
    U_0 = \exp\left({\rm i}\int_{0}^{t}H_{0}\ \mathrm{d}t\right),
\end{eqnarray}
we obtain the system Hamiltonian in the interaction picture%
\begin{eqnarray}
    H_{\rm I}^{\prime } = \varepsilon _{2}\cos (\nu _{2}t)\left\vert e\right\rangle \left\langle
    e\right\vert + \exp \left[ -{\rm i}\mu \sin (\nu _{1}t)\right] [-\lambda \exp
    \left( 2{\rm i}\nu _{1}t\right) a^{\dagger }+\Omega {\rm e}^{{\rm i}\theta }]\vert
    g\rangle \langle e\vert +\rm{H.c.} ,
\end{eqnarray}%
where $\mu =\varepsilon _{1}/\nu _{1}$. To clarify the underlying physics clearly, we use the Jacobi-Anger expansion
\begin{eqnarray}
\exp \left[ {\rm i}\mu \sin \left( \nu _{1}t\right) \right] =\stackrel{\infty }{%
    \mathrel{\mathop{\sum }\limits_{o=-\infty }}%
}J_{o}(\mu )\exp \left( {\rm i}o\nu _{1}t\right) ,
\end{eqnarray}
with $J_{o}(\mu )$ being the $o$th Bessel function of the first kind, we obtain
\begin{eqnarray}
    H_{\rm I}^{\prime } = \varepsilon _{2}\cos (\nu
    _{2}t)\vert e\rangle \langle e\vert +\stackrel{\infty }{%
        \mathrel{\mathop{\sum }\limits_{o=-\infty }}%
    }J_{o}(\mu )\left\{-\lambda \exp [-{\rm i}(o-2)\nu _{1}t]a^{\dagger }+\Omega {\rm e}^{{\rm i}\theta }\exp \left(
    -{\rm i}o\nu _{1}t\right) \right\}\vert g\rangle \langle e\vert
    +\rm{H.c.}.
\end{eqnarray}
For $\lambda$, $\Omega J_{0}(\mu )\ll \nu _{1}$, the qubit interacts with the resonator at the upper second-order sideband modulation with $o=2$, and is driven at the carrier\ with $o=0$. With the fast oscillating terms being discarded, $H_{\rm I}^{\prime }$ reduces to%
\begin{eqnarray}
H_{\rm I}^{\prime }=K\sigma _{\theta }+\frac{1}{2}\varepsilon _{2}\cos (\nu
_{2}t)\sigma _{z}-\eta {\rm e}^{-{\rm i}\theta }a^{\dagger }(\sigma _{\theta }-{\rm i}\sigma
_{\theta +{\rm \pi} /2})+\rm{H.c.},
\end{eqnarray}
where $K=\Omega J_{0}(\mu )$, $\eta =\lambda J_{2}(\mu )/2$, $\sigma_{\theta }={\rm e}^{{\rm i}\theta }\left\vert g\right\rangle \left\langle e\right\vert+{\rm e}^{-{\rm i}\theta }\left\vert e\right\rangle \left\langle g\right\vert $, and $\sigma _{z}=\left\vert e\right\rangle \left\langle e\right\vert -\left\vert g\right\rangle \left\langle g\right\vert $. With the assumption $\nu _{2}=2K\gg \eta$, $\varepsilon _{2}/2$ and under the further transformation $\exp ({\rm i}K\sigma _{\theta }t)$, the system Hamiltonian can be well approximated by
\begin{eqnarray}\label{eq10}
H_{\rm I}^{^{\prime \prime }}=-\eta {\rm e}^{-{\rm i}\theta }a^{\dagger }\sigma _{\theta
}+\rm{H.c.}+\omega \sigma _{z},
\end{eqnarray}
where $\omega =\varepsilon _{2}/4$. When $\theta ={\rm \pi} /2$, this effective Hamiltonian has the same form as the 1+1 Dirac equation of a spin-1/2 particle, with the correspondence $c^{\ast }=\sqrt{2}\eta $ and $m^{\ast}c^{\ast 2}=\omega $. Herein $c^{\ast }$ and $m^{\ast }$ denote the effective light speed and mass of the Dirac particle in the simulation, respectively.

\section*{Supplementary Note 2: System parameters}
The whole electronics and wiring of the used superconducting 5-qubit sample \cite{song2017,ning2019} are shown in Supplementary Figure~{wiring}.
Each frequency-tunable Xmon qubit has an individual flux line for its dynamic tuning and a microwave drive for controllably flipping its states. All the qubits are capacitively coupled to a bus resonator with coupling strength $\lambda_j\approx 2{\rm \pi}\times20$ MHz ($j=1,2$) and every qubit has a readout resonator for reading out its states. The bus resonator $R$ is a superconducting coplanar waveguide resonator with fixed frequency $\omega_{\rm r}/(2{\rm \pi}) = 5.5835$ GHz, which is measured when all used qubits stay in ground states at their respective idle frequencies (See below, unused qubits are always at their sweet points $\sim 6$ GHz). More detailed parameters of the sample are listed in Supplementary Table~\ref{par}.

In this experiment, we use two qubits ($Q_1, Q_2$) and the bus resonator ($R$). One qubit ($Q_1$) is used to realize the effective Dirac model, and the other qubit ($Q_2$) is used to measure the photon number distribution of the bus resonator by resonant population exchange (see Supplementary Note 5 for details).
For initialization, we wait for 300 $\mu$s to make sure all qubits return to their ground states, and then bias them to their idle frequencies, where single qubit gates and qubit states measurements are performed.

We use two independent XY signal channels controlled by Digital-to-Analog converters (DACs), to generate two microwave sequences. These two low-frequency microwave sequences are mixed with a continuous microwave by the In-phase and Quadrature (IQ) mixer to generate two tone pulses. This continuous microwave is a carrier microwave emitted by the microwave source (MS) and has a frequency of $5.21$ GHz. One-tone pulse is used to drive the test qubit to prepare the initial states, the other is used to trigger the resonator by crosstalk instead of the resonator line to achieve the displacement operation on its states in phase space, and to provide continuous driving pulses for Klein tunneling. For the Z signal, we use two independent channels on the DACs to control, which allows us to flexibly adjust the qubit frequency. In addition to the DACs described above, we need XY signal from the DACs to output the readout pulse with multiple tones implemented by sideband mixing. Then, the output pulse passes through the circulator, impedance-transformed Josephson parametric amplifier (JPA), cryogenic amplifier, and room temperature amplifier to improve signal-to-noise ratio. Finally, it is captured by Analog-to-Digital converters (ADCs) after passing through the readout resonator, and thus the multiple tones are demodulated to return the IQ information of each tone. The readout of the qubit states is performed at its idle frequency with the duration time of 1 $\mu$s. Both DACs and ADCs are supplied with the carrier microwave from the same MS. The frequency of the readout resonator $\omega_{ro} /(2{\rm \pi})$ ranges from 6.65 to 6.86 GHz, right in the bandwidth of our JPA with a pump frequency about $13.39$ GHz.
\begin{figure}[htb]
	\centering
	\includegraphics[width=14cm]{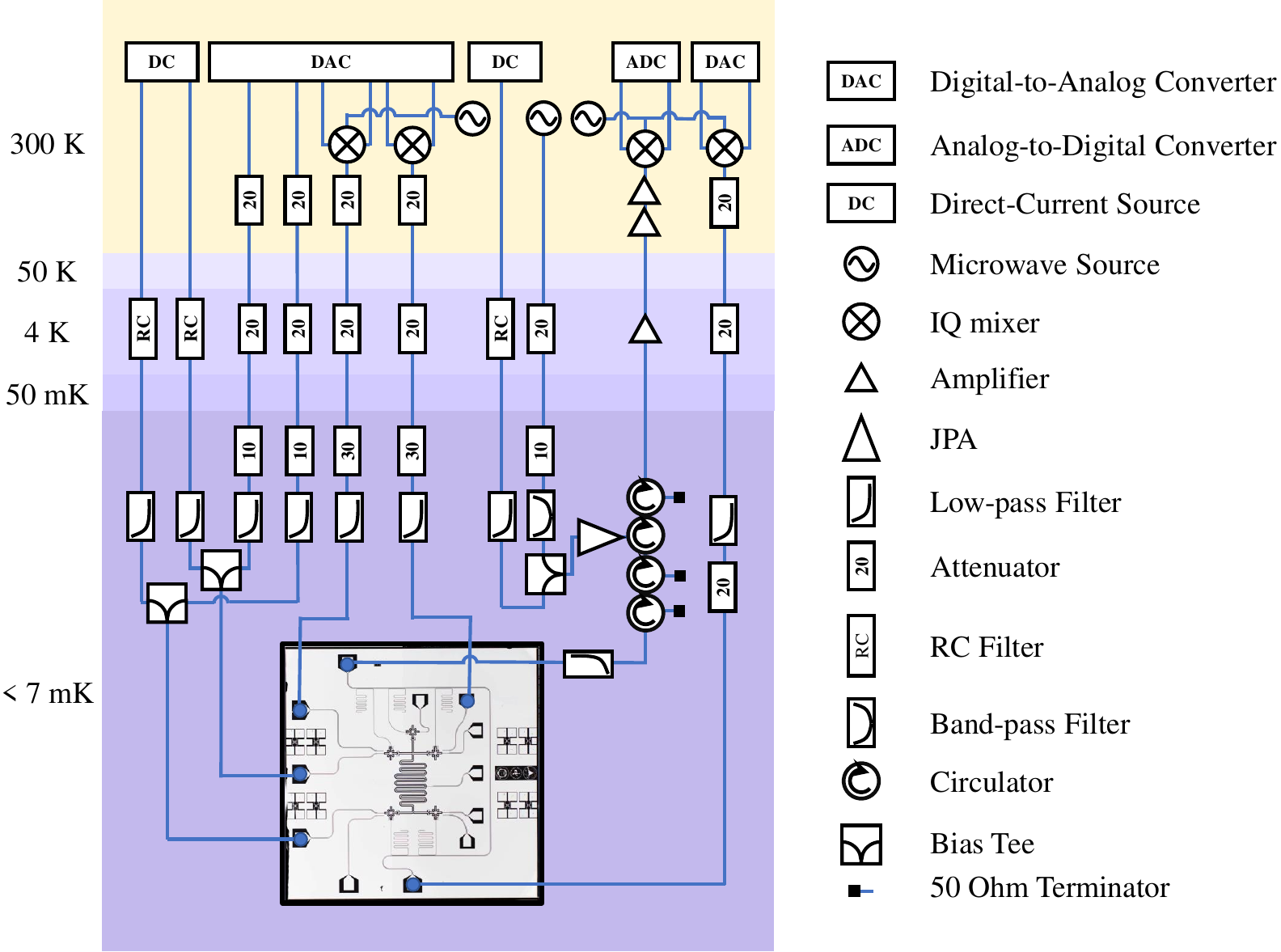}
	\caption{\textbf{Schematic diagram of the experimental setup.} Note there is a low-pass filter inverted with the others because it connects to the readout line. The number inside the attenuator icon indicates the magnitude of the attenuator in dBm.}
	\label{wiring}
\end{figure}
\begin{table}
	\centering
	\begin{tabular}{cccccccccc}
		\hline
		\hline
		& \ \ \multirow{2}{*}{$ \omega_{10}/(2{\rm \pi})$ (GHz)} \ \
		&\ \ \multirow{2}{*}{$T_{1}$ ($\mu$s)} \ \
		& \ \ \multirow{2}{*}{$T_{2}^*$ ($\mu$s)} \ \
		&\ \ \multirow{2}{*}{$T_{2}^{\textrm{SE}}$ ($\mu$s)}\ \
		& \ \ \multirow{2}{*}{$\lambda/(2{\rm \pi})$ (MHz)} \ \
		&\ \ \multirow{2}{*}{$\alpha/(2{\rm \pi})$ (MHz)} \ \
		&\ \ \multirow{2}{*}{$\omega_{ro}/(2{\rm \pi})$ (GHz)} \ \
		& \ \ \multirow{2}{*}{$F_{g}$} \ \ \ &\multirow{2}{*}{$F_{e}$}\\
		~\\
		\hline
		$Q_1$&5.260&21.5&1.1&6.0&19.91&250&6.764&0.983&0.937\\
		$Q_2$&5.110&17.2&1.5&14.3&20.92&238&6.655&0.990&0.920\\
		\hline
		$R$&5.5835&12.9&234.5&-&-&-&-&-&-\\
		\hline
		\hline
	\end{tabular}
	\caption{ \textbf{Qubit and resonator characteristics.}
		The test qubit, auxiliary qubit, and bus resonator are denoted by $Q_1, Q_2$, and $R$, respectively. The idle frequencies of $Q_j\ (j=1,2)$ are marked by $\omega_{10}/(2{\rm \pi})$, where the pulses for initial state preparation and tomographic pulses for qubit measurement are applied.
		The energy relaxation time, the Ramsey Gaussian dephasing time, and the spin echo Gaussian dephasing time are $T_1$, $T_2^*$ and $T_2^{\textrm{SE}}$, respectively, all measured at the idle frequency of each qubit.
		The coupling strength $\lambda_j$ between the qubits $Q_j$ and bus resonator $R$ is measured by their population exchange rate at resonance. The symbol $\alpha$ is the anharmonicity of the qubits, $\omega_{ro}$ is the bare frequency of qubit's readout resonator, and $F_g\ (F_e)$ is the probability of reading out $Q_j$ in $\vert g\rangle\ (\vert e\rangle)$ when it is prepared in $\vert g\rangle\ (\vert e\rangle)$.
	}
	\label{par}
\end{table}

During the procedure of qubit's initial state preparation and measurement, some errors are caused by decoherence. To calibrate these errors, we use a calibration matrix, defined as
\begin{eqnarray} \label{readcal}
	F_j = \left( \begin{matrix}
		F_{g,j} & 1-F_{e,j} \\ 1-F_{g,j} & F_{e,j}
	\end{matrix}\right),
\end{eqnarray}
where $F_{g,j}$ and $F_{e,j}$ are the readout fidelities of qubits $Q_j$ (see Supplementary Table~\ref{par}), to reconstruct the readout results. Defining $F = F_1 \otimes F_2$ as a two-qubit calibration matrix. We rewrite the measurement results of two qubits into a column vector $P_{\rm m}$, and the calibrated measurement result is then $P=F^{-1}\cdot P_{\rm m}$.


Here, the superconducting qubit can be approximated as a two-level system when the coupling $\lambda$ and transverse field strength $\Omega$ are relatively small (all $2{\rm \pi}\times20$ MHz) against the anharmonicity $2{\rm \pi}\times250$ MHz. The two-level system works well for simulating the internal degrees of freedom of a Dirac particle. On the other hand, the superconducting bus resonator, characterized by a quantized microwave field, was demonstrated \cite{Wallraff2004,Majer2017,Barends2014,Barends2013,song2017,Wang2008} to possess a bosonic mode, which can be utilized to emulate the external degrees of freedom of a Dirac particle. Overall, the present on-chip experimental platform can demonstrate a cavity QED system, with  advantages of adjustable artificial atomic energy spectrum and high coupling-to-decay ratio [$\lambda \gg (1/T_1, 1/T_2)$]. Nevertheless, current experiments are limited by the accuracy of measuring too many photons ($n\textgreater20$) in the microwave bosonic field. This limitation can be solved by increasing the detuning between the qubit and cavity ($\textgreater2{\rm \pi}\times800$ MHz), which requires the improvement of microwave signal control hardware in the present experimental platform.

\section*{Supplementary Note 3: Numerical Simulations of phase space Quantum Interference}

To further confirm the experimental results indeed reflect the quantum dynamics of Dirac particles, we perform numerical simulations based on the ideal Dirac Hamiltonian. The resonator WFs correlated with the qubit's $\vert g\rangle$ and $\vert e\rangle$ states after $330$ ns are respectively shown in Figs. \ref{sim_eff}(a) and \ref{sim_eff}(b), while the result irrespective of the qubit state is displayed in Supplementary Figure \ref{sim_eff}(c). In the numerical simulations, the Hamiltonian parameters and the system initial state are the same as those for the experimental simulations displayed in Fig. 2 of the main text. The numerical results with the full Hamiltonian are displayed in Supplementary Figure \ref{sim_full}. To compensate for the difference due to the Stark shifts produced by the discarded off-resonant couplings, we perform proper phase space rotations, which, however, do not change the quantum effects in any way. The numerical simulations of the evolutions of the mean position $\langle x \rangle $ and entropy are displayed in Supplementary Figure \ref{x_entropy}. All these numerical results agree well with the experimental simulations shown in Fig. 2 of the main text.

\begin{figure}[htb]
	\centering
	\includegraphics[width=12cm]{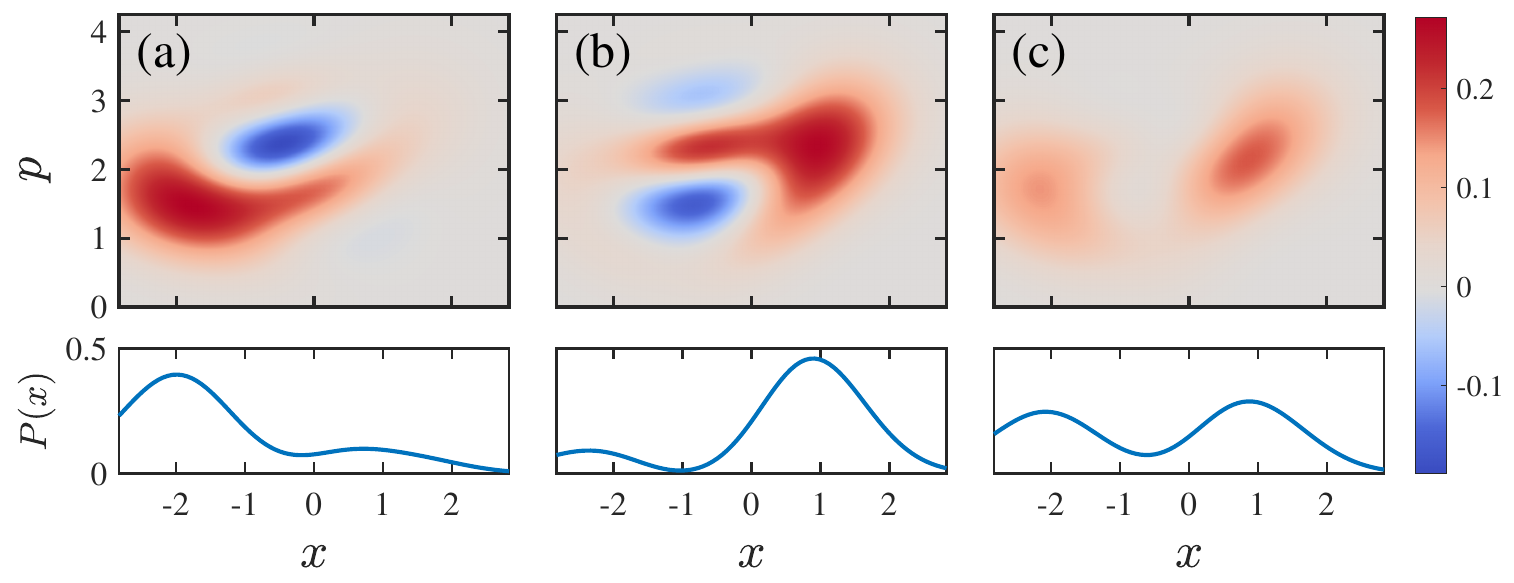}
	\caption{\textbf{Numerical simulations of phase space quantum interference, governed by the effective Hamiltonian Eq. (\ref{eq10}).}
	(a), (b) WFs correlated with the basis states $|g\rangle$ and $|e\rangle$ of the test qubit;
	(c) WFs irrespective of the test qubit’s state, all obtained after an evolution time of 330 ns.
	Lower panels: Probability distributions $P(x)$ with respect to the quadrature $x$, obtained by integrating the corresponding WFs over $p$. }
	\label{sim_eff}
\end{figure}

\begin{figure}[htb]
	\centering
	\includegraphics[width=12cm]{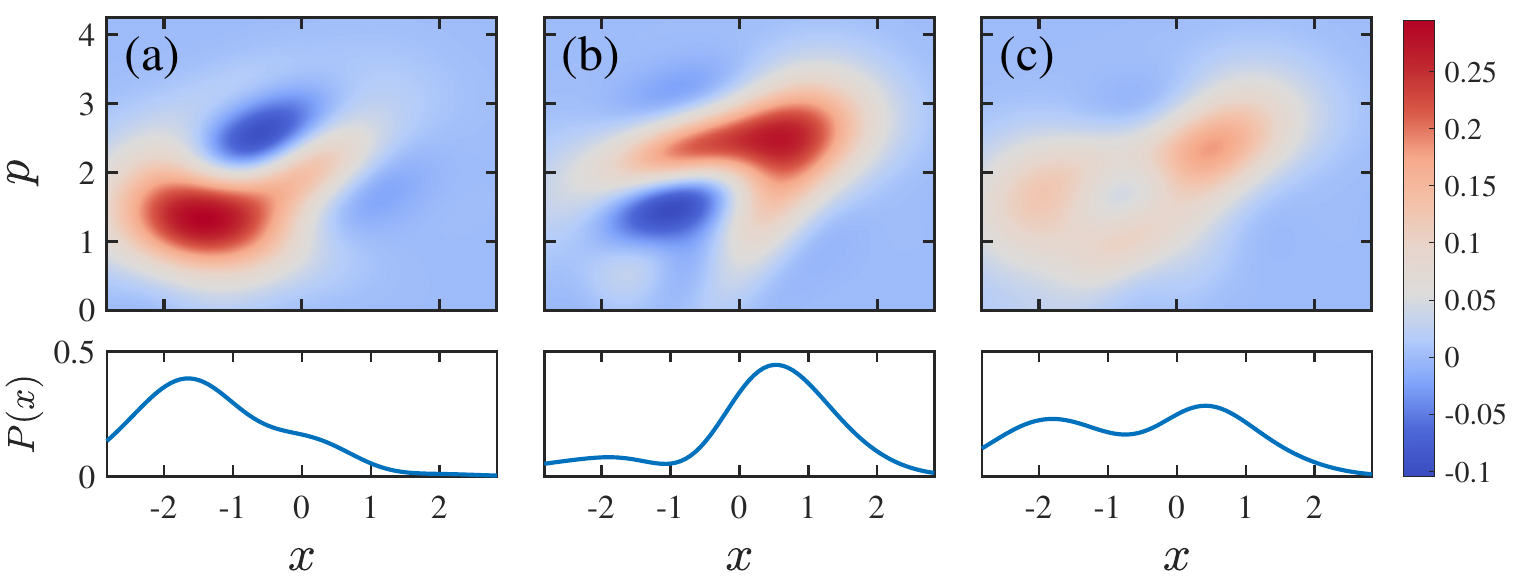}
	\caption{\textbf{Numerical simulations of phase space quantum interference, governed by the full Hamiltonian Eq. (\ref{full}).}
		(a), (b) WFs correlated with the basis states $|g\rangle$ and $|e\rangle$ of the test qubit;
		(c) WFs irrespective of the test qubit’s state, all deduced after an evolution time of 330 ns.
		Lower panels: Probability distributions $P(x)$ with respect to the quadrature $x$, obtained by integrating the corresponding WFs over $p$.}
	\label{sim_full}
\end{figure}

\begin{figure}[htb]
	\centering
	\includegraphics[width=12cm]{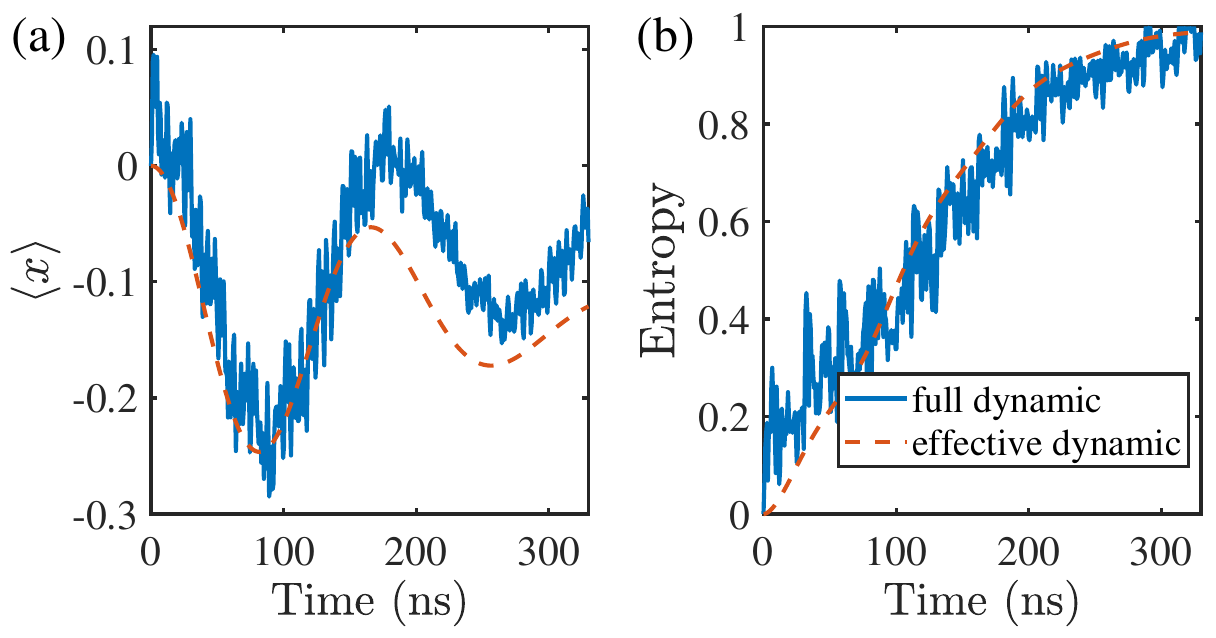}
	\caption{\textbf{Numerical simulation of {\sl Zitterbewegung}.} (a) Mean  position $\langle x\rangle $ versus time.
	(b) Entropy versus time.
	The solid and dashed lines represent the results of full and effective dynamics (governed by full and effective Hamiltonians), respectively.}
	\label{x_entropy}
\end{figure}
As stated in the main text, quantum interference is actually a universal inherent characteristic of Dirac particles, which can manifest itself even without negative components.
When the initial state is restricted to the positive branch, the system state can be written as
\begin{eqnarray}\label{pos}
|\psi_+(t)\rangle=\int {\rm d}p \  {\rm e}^{-{\rm i}E_pt} \xi_p |p\rangle |\phi_+(p)\rangle.
\end{eqnarray}
We assume that $|\xi_p|^2$ satisfies Gaussian distribution, i.e., $|\xi_p|^2=1/(\delta p \sqrt{2{\rm \pi}}) e^{-(p-p_0)^2/(2\delta p^2)}$, with $p_0=1$, $\delta p=1$.
To demonstrate the phase space quantum interference, we project the system state along the basis $\{| B_\pm \rangle\}=\{|e\rangle,|g\rangle\}$ and deduce the WF signal, given by
\begin{eqnarray}\label{pos_wig}
{\cal W}_{\pm}(x,p,t)=\frac{1}{{\rm \pi}}\int {\rm d}v \ \varphi_{\pm}^*(p+v)\varphi_{\pm}(p-v)  {\rm e}^{-2{\rm i}vx},
\end{eqnarray}
where $\varphi_{\pm}(p)=\xi_p^* {\rm e}^{-{\rm i}E_pt} \langle B_\pm  |\phi_+(p)\rangle$.
Numerical simulations of such WFs are shown in Supplementary Figure \ref{supp_pos_eigen_wig}, where we can see phase space quantum interference patterns for different evolution times.
The simulations are performed based on the ideal Dirac Hamiltonian with $m=1$, and by setting $\hbar=c=1$. An unexpected result is that the WFs, obtained irrespective of the internal state, also displays a time-evolving quantum interference pattern. This result is in distinct contrast with the case that the system is initially in the superposition of positive and negative components, for which the WFs exhibit a quantum interference pattern that appears only when correlating to the internal state. Such an unconditional feature further confirms the universality of the phase space quantum interference behavior.

Based on Eq. (\ref{pos}), the entropy is calculated by
\begin{eqnarray}\label{pos_entropy}
S'=-\log_2 \left(\int {\rm d}p \ |\xi_p|^2 |\phi_+(p)\rangle \langle \phi_+(p)| \right),
\end{eqnarray}
The numerical result is presented in Supplementary Figure \ref{pos_x_entropy}(a), which confirms that the internal and spatial degrees freedom remain highly correlated with a  time-independent entanglement entropy.
Additionally, the corresponding mean position $\langle x\rangle$ versus time is shown in Supplementary Figure \ref{pos_x_entropy}(b), which exhibits a linear increase without {\sl ZB}. These results unambiguously demonstrate that the Dirac particle evolves with a time-dependent quantum interference pattern, which is hidden in the entangled internal-spatial state, and expresses itself in the position-momentum quasiprobability distribution, correlated to the internal state.

\begin{figure}[htb]
	\centering
	\includegraphics[width=12cm]{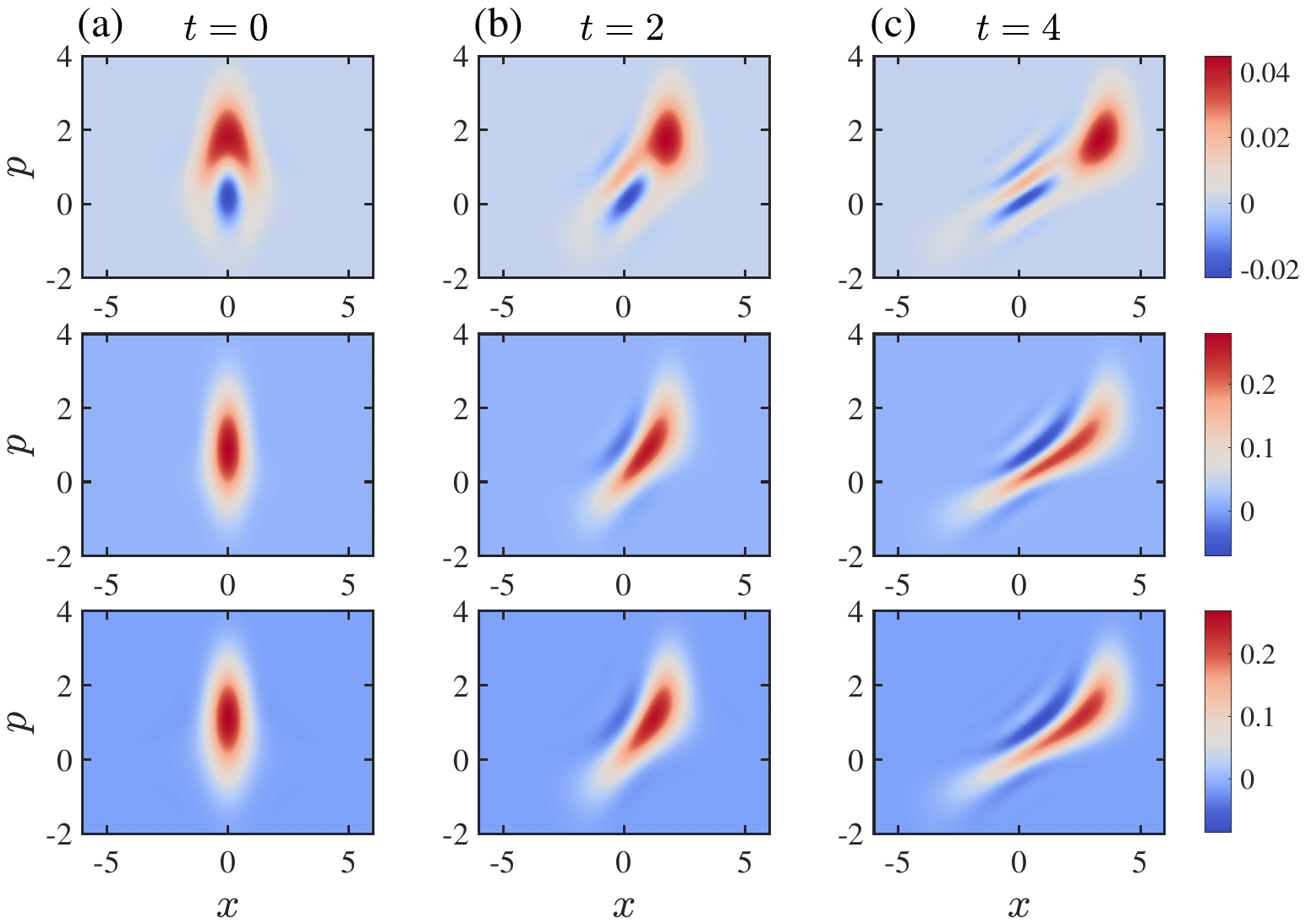}
	\caption{\textbf{Time-evolving WFs associated with the positive branch.}
	The upper and middle rows respectively denote the WFs correlated with the internal states $\vert g \rangle$ and $\vert e \rangle$, and the lower row shows the results irrespective of the internal state. The momentum is supposed to have a Gaussian distribution, centered at $p_0 =1$ with the spread $\delta p = 1$. The simulated Dirac particle has a mass of $m=1$. The results for $t=0$, 2, and 4 are shown in (a), (b), and (c), respectively. For simplicity, $\hbar$ and $c$ in the Dirac Hamiltonian are both set to be 1.}
	\label{supp_pos_eigen_wig}
\end{figure}
\begin{figure}[htb]
	\centering
	\includegraphics[width=10cm]{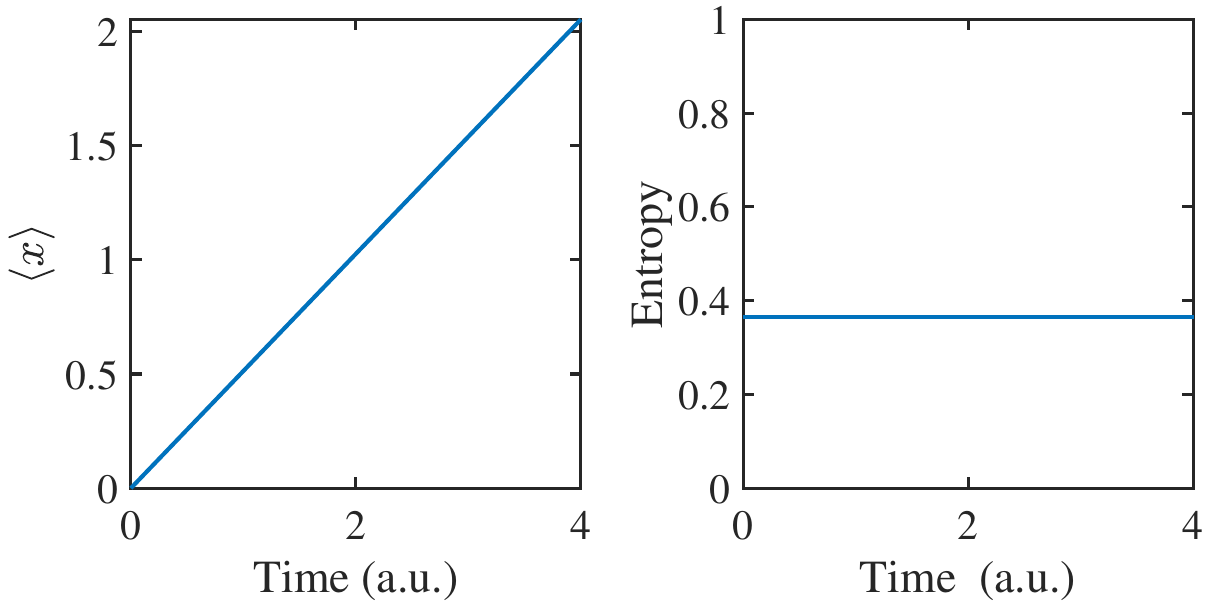}
	\caption{\textbf{Entanglement entropy evolution and spatial motion for the positive branch.}
	The mean position $\langle x\rangle$ is calculated by integrating $x{\cal W}(x,p,t)$ over phase space. }
	\label{pos_x_entropy}
\end{figure}

\section*{Supplementary Note 4: Optimization of driving pulses}\label{photon}
To achieve a periodic modulation of the qubit frequency, we first measure the resonant frequency $\omega_{10}$ versus the Z-line pulse amplitude (ZPA) for the test qubit, the results are shown in Supplementary Figure~\ref{freq2zpa} \cite{liu2020}. Here we set the detuning between the idle frequency of the $Q_1$ and the resonator frequency being $\Delta/(2{\rm \pi}) \approx 320$ MHz, and the frequency of the first modulation pulse $\nu_1/(2{\rm \pi}) = 160$ MHz. In such a case, the working frequency of the qubit is the idle frequency point, avoiding a large rising edge at the beginning of the pulse. Considering the frequency of the second frequency modulation pulse $\nu_2/(2{\rm \pi}) = 33.4$ MHz, this parameter setting will effectively reduce the high-energy level excitation caused by the modulation pulse. Subsequently, we modulate the qubit in the frequency with amplitude $\varepsilon_1 = 2{\rm \pi} \times 130$ MHz, centering around its idle frequency according to the $\omega_{q}$ versus ZPA curves obtained previously.

Considering that the nonlinear curve of frequency versus ZPA and the limited DACs bandwidth lead to imperfect waveforms of the periodically modulated excitation energy, we modify the frequency-modulated pulse as
\begin{equation}
    \omega_{\rm q}(t) = \omega_{0} + \delta + \varepsilon_1\cos(\nu_1t+\phi_1)+\varepsilon_2\cos(\nu_2t+\phi_2),
\end{equation}
making the experiment dynamics closer to the Dirac dynamics. We traverse these two phases ($\phi_1, \phi_2$) between 0 and $2{\rm \pi}$, set the initial state of the system to be $\vert \psi\rangle = \frac{1}{\sqrt2}(\vert e\rangle +\vert g\rangle )\otimes \vert 0\rangle$ in both the {\sl ZB} and Klein tunneling simulations, and finally obtain the one which is closest to the ideal situation. Here we use the 2-norm as an index to measure the agreement between the experimental population data of qubit state $\vert e \rangle$ and the theoretical values. Then, we optimize the $\delta$ and other parameters in the same way. In this case, the optimized population is shown in Supplementary Figure~\ref{rabicheck}, and the experimental data are in good agreement with the ideal ones, which proves the correctness of our method.

For the transverse field driving the qubit, we want the microwave to arrive at the qubit with an initial phase ${\rm \pi}/2$. Therefore, we use a similar approach to optimize this phase. In addition, in the application of modulated pulses, the oscillation center frequency of the qubit may deviate slightly from the working frequency $\omega_0$, so we mildly tune the microwave frequency (about $1\sim 2$ MHz) of the transverse field to make the results more predictable.
\begin{figure}[htb]
    \centering
    \includegraphics[width=8cm]{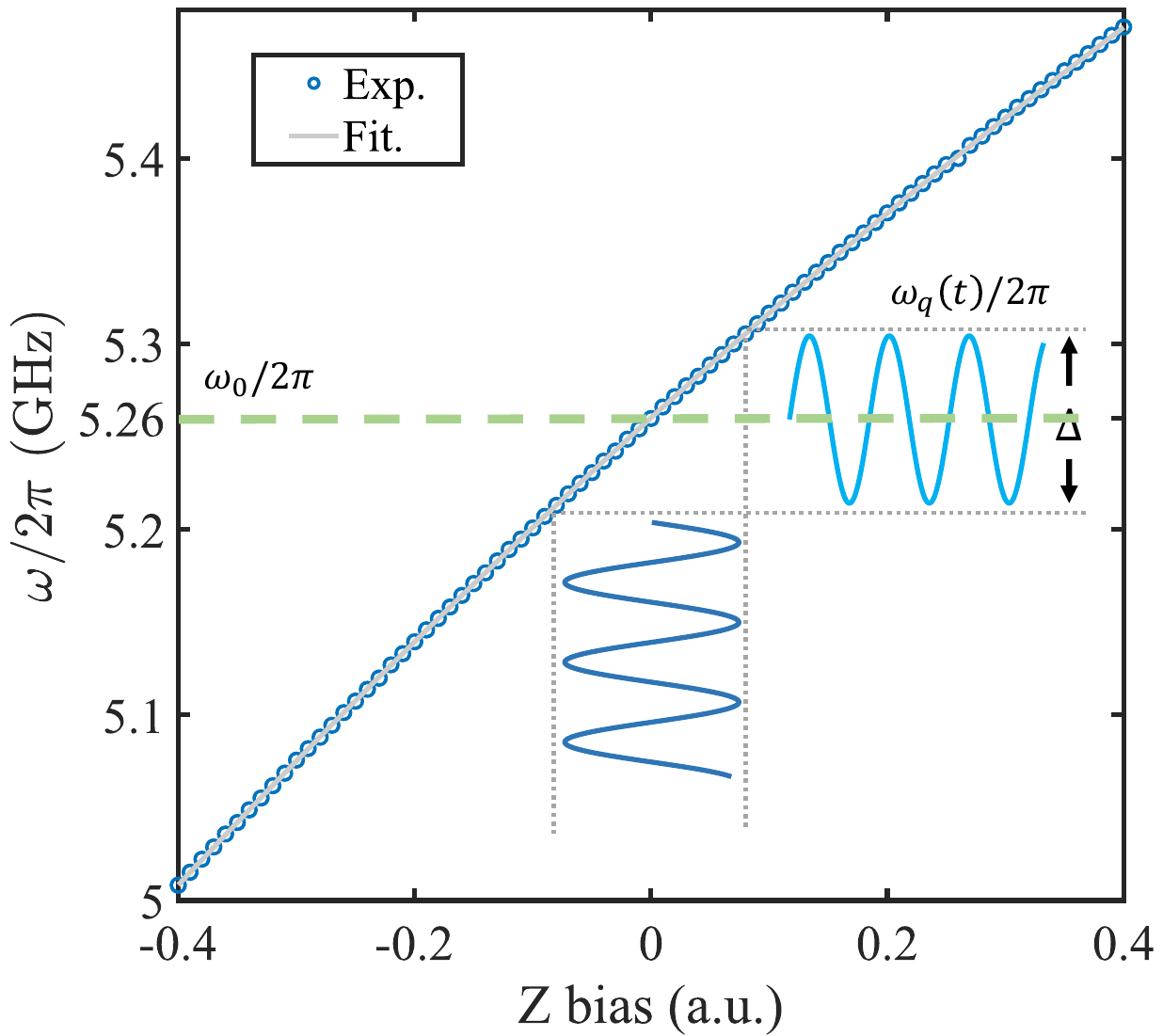}
    \caption{\textbf{Frequence of the test qubit versus Z line bias.} The desired modulation of $\omega_{q}(t)$ can be realized by mapping it to the modulation of Z bias.}
    \label{freq2zpa}
\end{figure}

\begin{figure}
	\includegraphics[width=12cm]{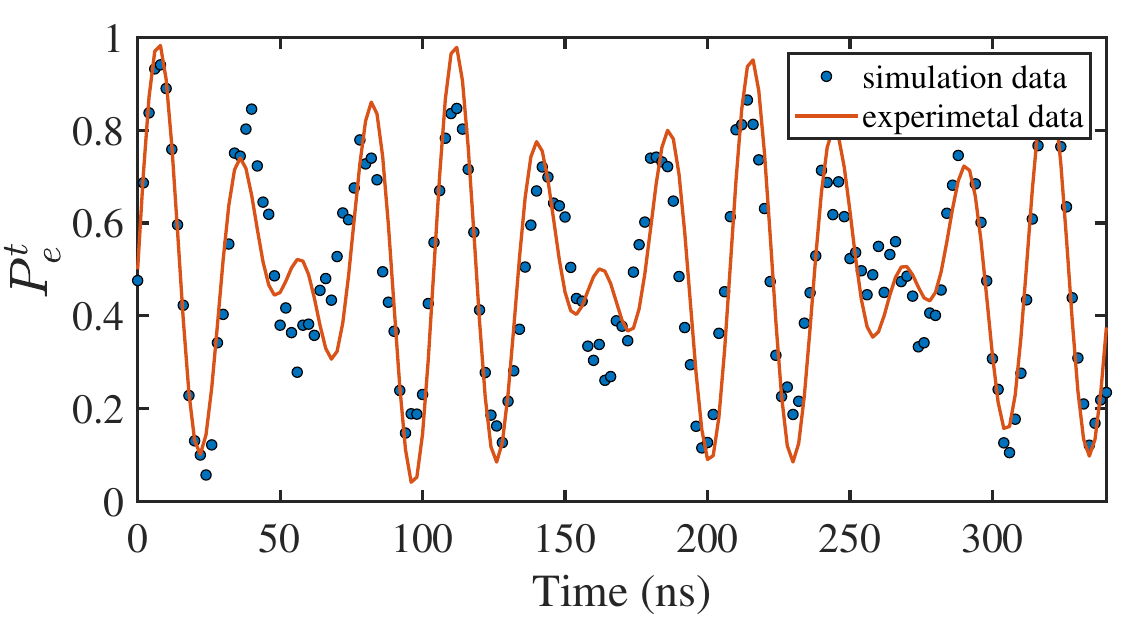}
	\caption{\textbf{Validity check of the Rabi model.}
		For the present Rabi model, the effective frequency of the qubit and the effective coupling strength are $\omega/(2{\rm \pi})=2.2$ MHz and $\eta=2{\rm \pi}\times$ 0.78 MHz, respectively.
	    We fix the initial state as $\vert \psi_0\rangle = \frac{1}{\sqrt2}(\vert g\rangle +\vert e\rangle )\otimes \vert 0\rangle$.
		After optimizing parameters $\varepsilon_1$, $\varepsilon_2$, $\phi_1$, $\phi_2$, and $\delta$ appropriately, in the experiment, we measure the population of the test qubit state $|e\rangle$, marked as $P_e^t$, at different times, and compare it with the numerical results.
		}
	\label{rabicheck}
\end{figure}

The resonator is driven by using a flattop envelope with a resonance frequency, described as $\Omega_{\rm r} (a+a^\dagger)$, where $\Omega_{\rm r}$ is the Rabi frequency of the pulse. During the preparation of the initial state in the {\sl ZB} simulation experiment, the presence of a dispersion coupling (form of $\sigma_{z} a^\dagger a$) between the test qubit and the bus resonator is $\lambda_1 ^2/\Delta\approx 2{\rm \pi}\times 1.25$ MHz. So we reduce the pulse time and increase the Rabi frequency of the pulse to avoid the initial state rotation caused by the dispersion interaction. Lengths of pulse for initial state preparation and tomography pulse are 24 ns and 60 ns, respectively.

For pulses of a specific length, we apply them with different amplitudes to the resonator and measure the photon number to fit the slope between the Rabi frequency ($\Omega_r$) and the pulse amplitude. Up to now, together with the initial phase of the microwave pulse, we can implement arbitrary displacement operators $D(\gamma)$ for Wigner measurements.

The Klein tunneling experiment requires a continuous pulse on the resonator during the dynamics. There is a fixed resonator frequency shift induced by the non-resonant coupling, thus we adjust the microwave frequency to about $0.75$ MHz to get better results.

The qubit-state-dependent resonator frequency shift in the dynamical process is still unavoidable; therefore, additional correction of the results is necessary so as to make the results more intuitive. (see Supplementary Note 6).

The pulse sequence is shown in Supplementary Figure~\ref{pulseseq}, including three steps: 1. Initial state preparation; 2. Dirac dynamics; 3. Quantum state measurement. For clarity of reading, the real-time scales are not used in the figure.

\begin{figure}[htb]
    \centering
    \includegraphics[width=14cm]{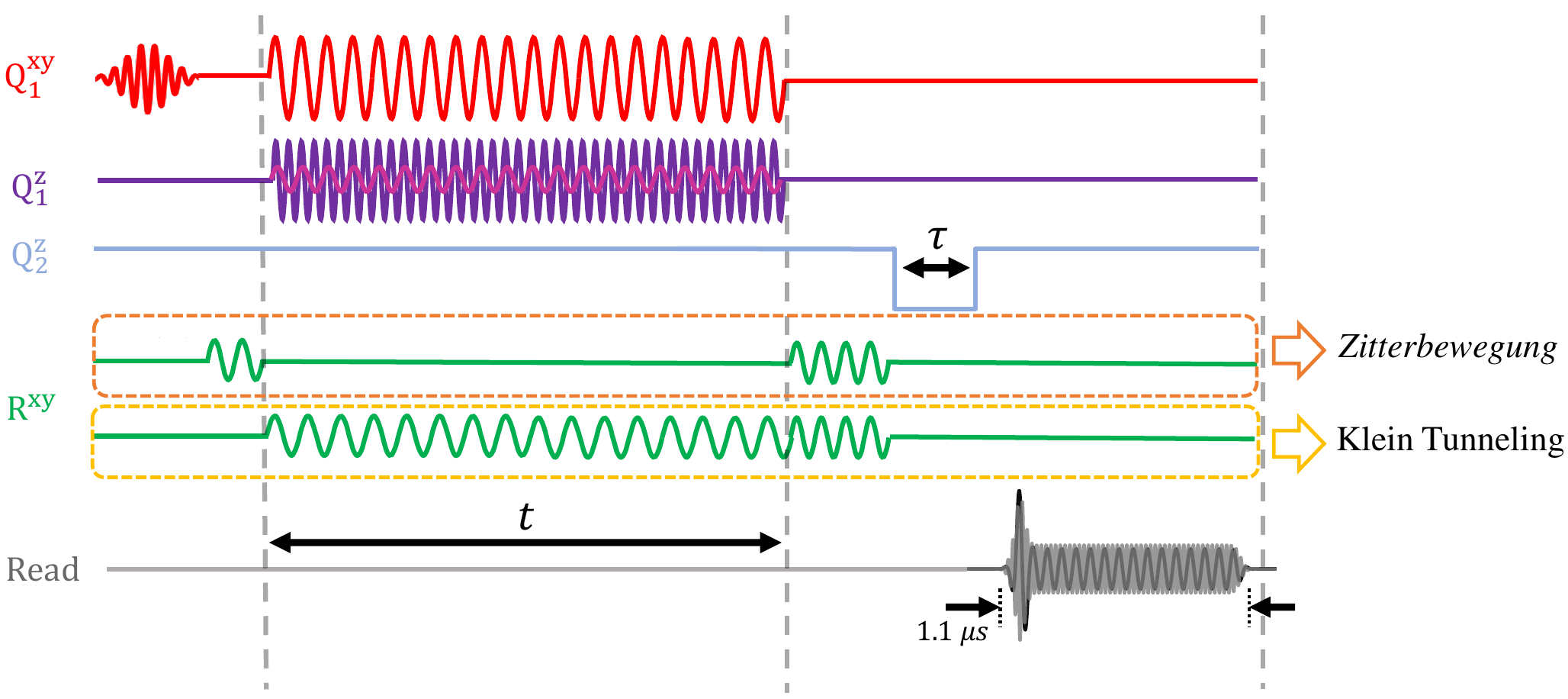}
    \caption{
        \textbf{Sketch of the Pulse Sequences.}
        The test qubit $Q_1$ is driven by a transverse microwave field with the Rabi frequency $\Omega/(2{\rm \pi})\approx20.03$ MHz through the XY-line (red), and two Sine longitudinal modulations apply to the Z-line (purple) with the modulation frequencies $\nu_{1(2)}/(2{\rm \pi})=160\ (33.4)$ MHz and the amplitudes $\varepsilon_{1(2)}=2{\rm \pi} \times 130\  (8.8)$ MHz in the {\sl ZB} simulation. While in the Klein tunneling simulation, we set $\varepsilon_2 = 0$ to simulate massless particles.
        The orange dashed box represents the pulse sequence of $\mathrm{R^{xy}}$ used in the {\sl ZB} simulation, which is replaced by the series of yellow dashed boxes in the Klein tunneling simulation, and outside the dashed box there are the same pulse sequences for these two simulations. The evolution times $t$ of the two experiments are 330 ns and 288 ns, respectively.
        The auxiliary qubit $Q_2$ is used to measure the resonator photon number. After the displacement operation pulse, we bias $Q_2$ to the resonator frequency for a given time $\tau$ and then deduce the photon number distribution of the resonator according to the result of the Rabi oscillation.
        Finally, a multiplexing tone is applied to the readin line to simultaneously measure two qubits' states.
    }
    \label{pulseseq}
\end{figure}
\section*{Supplementary Note 5: Characterization of the Qubit-Resonator State}
All the WF values in the main text are deduced from the photon number distribution. We use the same method as in Ref. \cite{zheng2022}. After the Dirac dynamics and displacement operation, we bias the auxiliary qubit $Q_2$ to the frequency of the bus resonator for a given time $\tau$. Then, we bias it to the idle frequency and measure. The excited state population $P^a_e(\tau)$ of $Q_2$ is defined as
\begin{eqnarray}\label{Pnfit}
    P_{e}^{a}(\tau)=\frac{1}{2}\left[1-P_g^a(0)\stackrel{n_{\max }}{%
        \mathrel{\mathop{\sum }\limits_{n=0}}%
    }P_{n}{\rm e}^{-\kappa _{n}\tau}\cos \left(2\sqrt{n}\lambda_2 \tau\right)\right],
\end{eqnarray}
where $P_{n}$ denotes the photon-number distribution probability, $n_{\max} $ is the cutoff of the photon number, and $\kappa_{n}=n^{l}/T_{1,p}$ ($l=1$) \cite{Pnfit1,Pnfit2,Pnfit3,Pnfit4,Pnfit5} is the empirical decay rate of the $n$-photon state, $\lambda_2$ is the coupling strength between auxiliary qubit $Q_2$ and resonator. It is worth noting that due to the finite detuning $\Delta = 2{\rm \pi}\times470$ MHz between the auxiliary qubit $Q_2$ and bus resonator $R$, as the number of photons increases, ancilla qubit $Q_2$ inevitably interacts with the resonator during the Dirac dynamics, the ground state population $P^a_g(0)$ may not be 1 at $\tau=0$. Referring to \cite{Pnfit1}, the potentially slight excitation can be ignored and thus $P^a_g(0)$ is introduced to make the fit more accurate. In the experiment, $P^a_g(0) \leq 0.05$, $\tau$ is taken every 2 ns from $0$ to $200$ ns. We use the least square method to fit the experimental data according to Eq.~(\ref{Pnfit}) to obtain the actual photon number distribution and further obtain the value of the WF. However, with the further increase of the photon number, the entanglement between the auxiliary qubit $Q_2$ and the resonator can not be ignored, and the photon number and its distribution in the resonator can not be obtained accurately. It is also worth noting that although ancilla qubits can be biased further down to increase the detuning in the Dirac dynamics, the performance of the qubits deteriorates and the frequency changes more widely, affecting the measurement results. Ergo, the photon number capability of the present system is about 20, resulting in the boundary of phase space in the main text.

The WF is given by
\begin{eqnarray}\label{wf}
        {\cal W}(x,p)&=&\frac{1}{{\rm \pi} }\stackrel{\infty }{\mathrel{\mathop{\sum }\limits_{n=0}}}(-1)^{n}{\cal P}_{n}(\gamma),\\
    {\cal P}_{n}(\gamma)&=&\left\langle n\right\vert D(-\gamma )\rho D(\gamma )\left\vert n\right\rangle,
\end{eqnarray}
where $x=\sqrt{2}\text{Re}(\gamma)$, $p=\sqrt2\text{Im}(\gamma)$, $D(\gamma )=\exp \left(\gamma a^{\dagger }-\gamma^{*}a\right)$. The photon number distribution ${\cal P}_{n}$ is inferred by the Rabi oscillation signal \cite{Pnfit5}. And then we calculate the Wigner matrix according to Eq.~(\ref{wf}). We adjust the Wigner tomography pulse based on $\gamma$ to realize the measurement of WF values at different positions in phase space.
The WF conditional on test qubit states $\vert e\rangle $ and $\vert g\rangle $ are given by
\begin{eqnarray}
    {\cal W}_{e(g)}(x,p)&=&\frac{1}{{\rm \pi} }\stackrel{\infty }{\mathrel{\mathop{\sum }\limits_{n=0}}}(-1)^{n}{\cal P}^{e(g)}_{n}(\gamma),\\
    {\cal P}^{e(g)}_{n}(\gamma)&=&\frac{1}{P_k^t}\left\langle n\right\vert \otimes \left\langle e(g) \right\vert D(-\gamma )\rho D(\gamma )\left\vert e(g)\right\rangle\otimes\left\vert n\right\rangle,
\end{eqnarray}
with  $P_k^t$ being the population of the test qubit $Q_1$ in $\vert k\rangle$ state at time $t$.

In this way, we can get all the information about the resonator state. Then, we use the CVX toolbox based on MATLAB \cite{cvx} to reconstruct the corresponding density matrix $\rho_k$ to calculate the average position $\langle x\rangle $ and average momentum $\langle p\rangle $ of the resonator.

As mentioned in the main text, a larger number of photons requires Wigner tomography within a larger phase space region. Due to the hardware-limited pulse amplitude, we can increase the size of $\vert\gamma\vert$ by increasing the length of the tomography pulse, at the cost of increasing the dispersive interaction time. Considering the finite detuning between the auxiliary qubit and the resonator, our {\sl ZB} simulation ends at 330 ns. With the improvement of the hardware, such as larger pulse amplitude, or better qubits performance, we could simulate the {\sl ZB} behavior within a longer time scale.
\section*{Supplementary Note 6: Correction of phase space Rotations due to Stark Shifts}\label{wfrota}
After reconstructing the density matrix $\rho_k$ from the Wigner tomography ${\cal W}_k$, we add a rotation operator on the corresponding density matrix numerically. The rotation operator is defined as
\begin{eqnarray}\label{rotation}
U_k&=&\exp\left[-{\rm i}(\theta_{k,0} + \theta_k t/t_f) a^\dagger a\right], \\
\rho_{k}^{\prime} &=& U_k \rho_k U_k^\dagger,\ \ \ \ \ \ \ \ \ \ (k=e,g)
\end{eqnarray}
where $\theta_{k,0}$ is used to counteract the resonator states rotation due to the initial state preparation when the test qubit in $\vert k \rangle$ state, $\theta_k$ cancels out the rotation caused by the difference between the experimental frame and the effective Hamiltonian frame in Eq.~(\ref{eq10}), $t_f$ is the end time of each simulation, and $\rho^{\prime}_{k}$ is the resonator density matrix after rotation when the test qubit in $\vert k\rangle$ state. In the Klein tunneling simulation experiment, because the initial state of the resonator is a vacuum state, we set $\theta_{k,0} = 0$.
Here we choose $\left(\theta_{e,0},\theta_{g,0}, \theta_e, \theta_g, t_f\right) = \left(0.260, 0.045, 1.510, 1.217, 330 \ \mathrm{ns} \right)$ in the {\sl ZB} simulution and $\left(\theta_{e,0},\theta_{g,0},\theta_e, \theta_g, t_f\right) = \left( 0, 0, 1.402, 1.162, 280 \ \mathrm{ns}\right)$ in the Klein tunneling simulation.
Note $\theta_e$ and $\theta_g$ are slightly different because the dispersive interaction during Wigner tomography (time scale 60 ns, causing phase difference $2 \lambda_1^2/ \Delta \times 60\ \mathrm{ns} \sim 0.3$). While both $\theta_e$ and $\theta_g$ are around $1.3$, offsetting the resonator phase induced by the frame difference, and therefore demonstrating that the rotation of our data is reasonable.
The unconditional resonator WF is obtained by adding two rotated WFs according to the corresponding qubits state population:
\begin{eqnarray}\label{wigreson}
    {\cal W}^{\prime}(t) = P_e^t{\cal W}^{\prime}_e(t) + P_g^t{\cal W}^{\prime}_g(t),
\end{eqnarray}
where ${\cal W}^{\prime}_k(t)$ is calculated from $\rho^{\prime}_{k}$. The complete Wigner data are shown in Supplementary Figure~\ref{ZB_wigdata} for {\sl ZB} simulation and Supplementary Figure~\ref{Klein_wigdata} for Klein tunneling simulation. For the selection of the time point, we try to make the population of states $\vert e\rangle $ and $\vert g\rangle $ of the test qubit close to $0.5$, making the normalized data of the auxiliary qubit accurate and convenient to fit the photon number distribution of the resonator.


\begin{figure}[htb]
	\includegraphics[width=16cm]{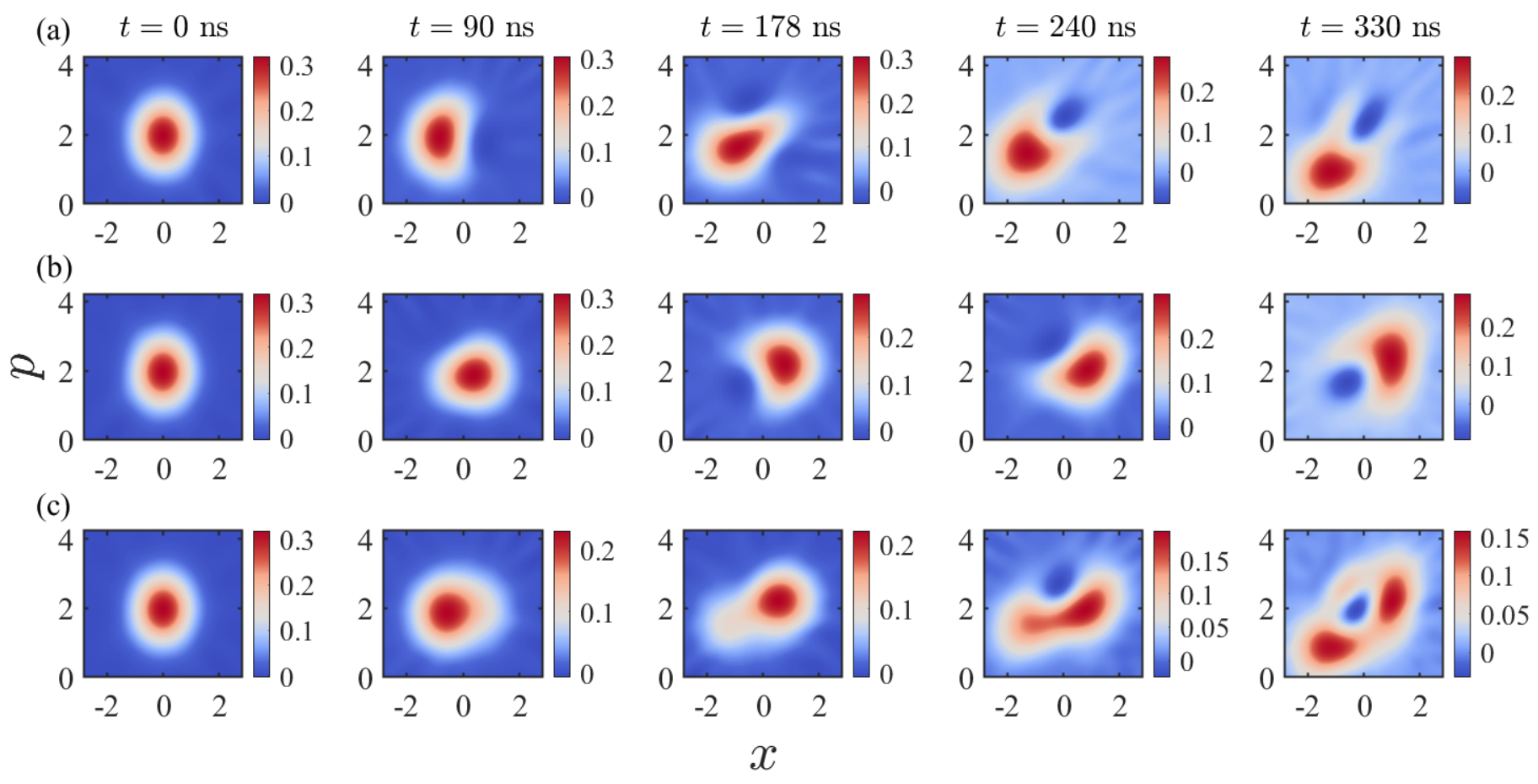}
	\caption{\textbf{Wigner functions for {\sl Zitterbewegung}.}
	The WFs correlated with the basis states $|g\rangle$ and $|e\rangle$ of the test qubit displayed in (a) and (b), respectively, and (c) show the result irrespective of the test qubit’s state.
	These five columns show the WFs at different times: 0, 90, 178, 240, and 330 ns.
	}
	\label{ZB_wigdata}
\end{figure}
\begin{figure}[htb]
	\includegraphics[width=16cm]{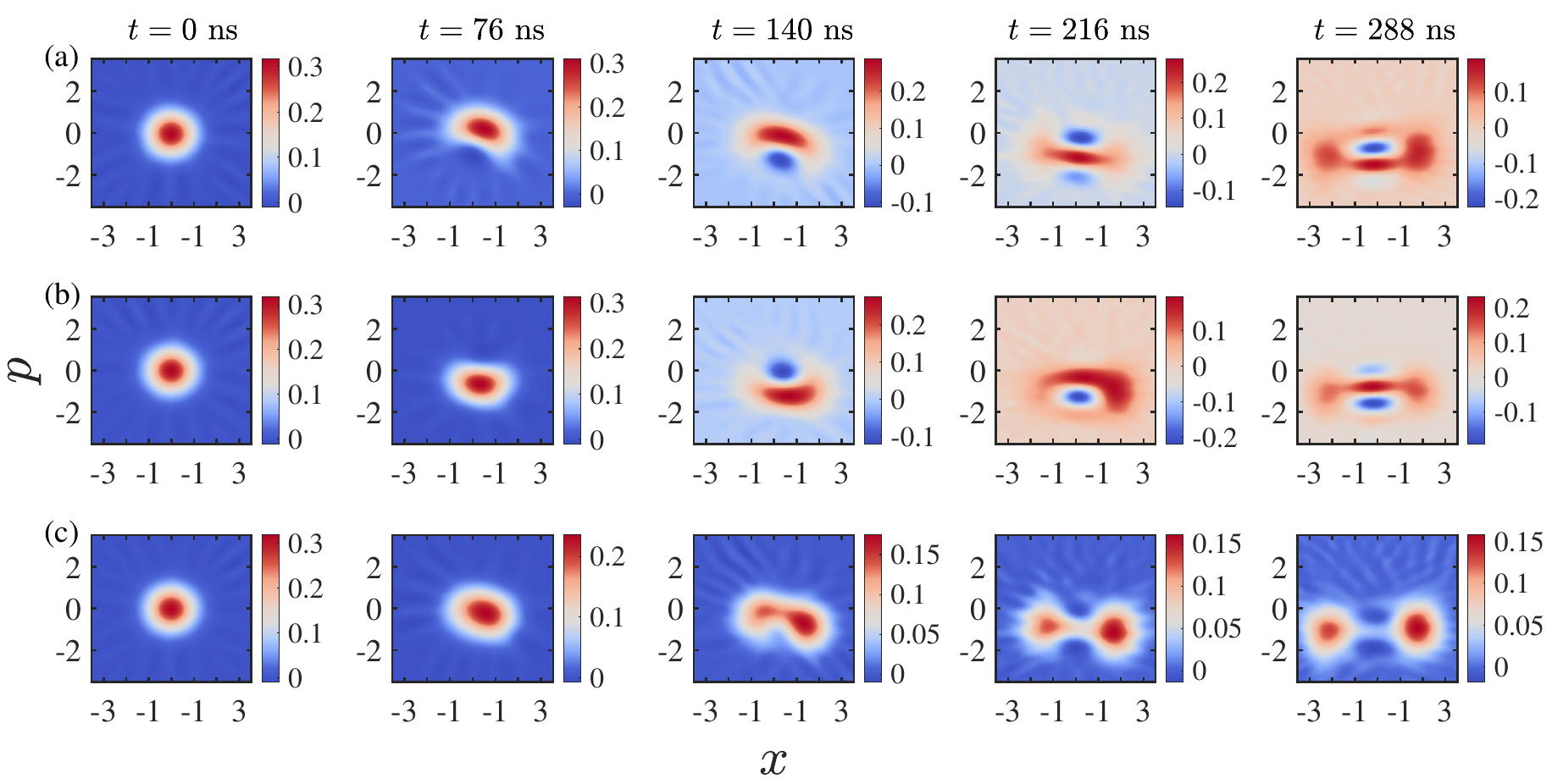}
	\caption{\textbf{Wigner function for Klein tunneling.}
		The WFs correlated with the basis states $|g\rangle$ and $|e\rangle$ of the test qubit displayed in (a) and (b), respectively, and (c) show the result irrespective of the test qubit’s state.
		These five columns show the WFs at different times: 0, 76, 140, 216, and 288 ns.}
	\label{Klein_wigdata}
\end{figure}
\section*{Supplementary Note 7: Discrimination between Positive and Negative components in Klein Tunneling}
In the Klein tunneling simulation, we need to show the measured position average $\langle x \rangle$ and momentum average $\langle p \rangle$ for the positive and negative wavepackets. Here, we simply use $x=0$ as a boundary to distinguish positive and negative wavepackets. The measured $\langle x \rangle $ and $\langle p \rangle $ for positive wavepackets can be calculated by
\begin{eqnarray}\label{meanx}
    \langle x\rangle &=&\frac{\int_{-\infty}^{{+\infty}} \int_{0}^{{+\infty}}  x{\cal W}(x,p)\  \mathrm{d}x\mathrm{d}p}{\int_{-\infty}^{{+\infty}} \int_{0}^{{+\infty}} {\cal W}(x,p)\ \mathrm{d}x\mathrm{d}p},\\
    \langle p\rangle &=&\frac{\int_{-\infty}^{{+\infty}} \int_{0}^{{+\infty}}  p{\cal W}(x,p)\  \mathrm{d}x\mathrm{d}p}{\int_{-\infty}^{{+\infty}} \int_{0}^{{+\infty}} {\cal W}(x,p)\ \mathrm{d}x\mathrm{d}p},
\end{eqnarray}
where ${\cal W}(x,p)$ is shown in Supplementary Figure~\ref{Klein_wigdata}(c). At $t=0$ and $76$ ns, the positive and negative wavepackets cannot be fully distinguished, so the data for these two mean positions $\langle x \rangle$ are abandoned in Fig. 3(d) of the main text. The negative wavepacket can be obtained in the same way,
\begin{eqnarray}\label{meanx2}
    \langle x\rangle &=&\frac{\int_{-\infty}^{{+\infty}} \int_{-\infty}^{{0}}  x{\cal W}(x,p) \ \mathrm{d}x\mathrm{d}p}{\int_{-\infty}^{{+\infty}} \int_{-\infty}^{{0}} {\cal W}(x,p)\ \mathrm{d}x\mathrm{d}p},\\
    \langle p\rangle &=&\frac{\int_{-\infty}^{{+\infty}} \int_{-\infty}^{{0}}  p{\cal W}(x,p) \ \mathrm{d}x\mathrm{d}p}{\int_{-\infty}^{{+\infty}} \int_{-\infty}^{{0}} {\cal W}(x,p)\ \mathrm{d}x\mathrm{d}p}.
\end{eqnarray}
\section*{SUPPLEMENTARY REFERENCES}